\documentclass[twocolumn,tighten,trackchanges]{aastex62}
\usepackage{graphicx}		% needed for the figures
\usepackage{url}
\usepackage{epstopdf}
\usepackage{color}
\usepackage{multirow}
\usepackage{ragged2e}
\usepackage{hyperref}
\usepackage{url}
\DeclareGraphicsExtensions{.pdf,.png,.jpeg}
\def\sgra{Sgr\,A$^{\ast}$}
\def\lsim{\mathrel{\raise.3ex\hbox{$<$\kern-.75em\lower1ex\hbox{$\sim$}}}}
\def\gsim{\mathrel{\raise.3ex\hbox{$>$\kern-.75em\lower1ex\hbox{$\sim$}}}}
\def\gtwid{\mathrel{\raise.3ex\hbox{$>$\kern-.75em\lower1ex\hbox{$\sim$}}}}
\def\proptwid{\mathrel{\raise.3ex\hbox{$\propto$\kern-.75em\lower1ex\hbox{$\sim$}}}}

\defcitealias{Johnson_2018}{J18}
\defcitealias{Brinkerink_2018}{B18}
\defcitealias{Brinkerink_2016}{B16}
\defcitealias{Ortiz_2016}{O16}
\defcitealias{Goldreich_2006}{GS06}

\def\apjl{ApJL}%           % Astrophysical Journal, Letters 
\def\apjs{ApJS}%           % Astrophysical Journal, Supplement 
\def\ao{Appl.~Opt.}%           % Applied Optics 
\def\aap{A\&A}%           % Astronomy and Astrophysics 
\newcommand{\nrao}{National Radio Astronomy Observatory, 520 Edgemont Rd, Charlottesville, VA, 22903, USA}
\newcommand{\naoj}{National Astronomical Observatory of Japan, 2-21-1 Osawa, Mitaka, Tokyo 181-8588, Japan}
\newcommand{\haystack}{Massachusetts Institute of Technology, Haystack Observatory, 99 Millstone Rd, Westford, MA 01886, USA}

\newcommand{\perimeter}{Perimeter Institute for Theoretical Physics, 31 Caroline Street, North Waterloo, Ontario N2L 2Y5, Canada}
\newcommand{\asiaa}{Institute of Astronomy and Astrophysics, Academia Sinica, P.O. Box 23-141, Taipei 10617, Taiwan}
\newcommand{\radboud}{Department of Astrophysics/IMAPP, Radboud University, P.O. Box 9010, 6500 GL Nijmegen, The Netherlands}

\newcommand{\cfa}{Harvard-Smithsonian Center for Astrophysics, 60 Garden Street, Cambridge, MA 02138, USA}
\newcommand{\mpifr}{Max-Planck-Institut f{\"u}r Radioastronomie, Auf dem H{\"u}gel 69, D-53121 Bonn, Germany}

\newcommand{\leiden}{ALLEGRO/Leiden Observatory, Leiden University, PO Box 9513, 2300 RA, Leiden, The Netherlands}
\newcommand{\chalmers}{Department of Space, Earth and Environment, Chalmers University of Technology, Onsala Space Observatory, 439 92 Onsala,
Sweden}
\newcommand{\mpe}{Max-Planck-Institut f{\"u}r Extraterrestrische Physik, Giessenbachstr. 1, 85748 Garching, Germany}
\newcommand{\frankfurt}{Institut f{\"u}r Theoretische Physik, Johann Wolfgang Goethe-Universit{\"a}t, Max-von-Laue-Stra{\ss}e 1, 60438 Frankfurt, Germany}
\newcommand{\arizona}{University of Arizona, 933 North Cherry Avenue, Tucson, AZ 85721, USA}
\newcommand{\granada}{Instituto de Astrof{\'i}sica de Andaluc{\'i}a-CSIC, Glorieta de la Astronom{\'i}a s/n, E-18008 Granada, Spain}
\newcommand{\kasi}{Korea Astronomy and Space Science Institute, Daedeok−daero 776, Yuseong−gu, Daejeon 34055, Korea}
\newcommand{\api}{Anton Pannekoek Institute for Astronomy, University of Amsterdam, 1098 XH Amsterdam, The Netherlands}
\newcommand{\yebes}{Observatorio de Yebes (IGN), Apartado 148, 19180, Yebes, Spain}
\newcommand{\iram}{Institut de RadioAstronomie Millim{\'e}trique (IRAM), Granada, Spain}
\newcommand{\shanghai}{Shanghai Astronomical Observatory, Chinese Academy of Sciences, Shanghai 200030, China}
\newcommand{\unam}{Instituto de Radioastronom{\'i}a y Astrof{\'i}sica, Universidad Nacional Aut{\'o}noma de Mexico, Morelia 58089, M{\'e}xico }
\newcommand{\unamm}{Instituto de Astronom{\'i}a, Universidad Nacional Aut{\'o}noma de M{\'e}xico, Apartado Postal 70-264, 04510 Ciudad de M{\'e}xico, M{\'e}xico }
\newcommand{\ust}{University of Science and Technology, Gajeong-ro 217, Yuseong-gu, Daejeon 34113, Korea}
\newcommand{\irap}{IRAP, Université de Toulouse, CNRS, UPS, CNES, Toulouse, France}

\begin{document}

\title{ 
The Size, Shape, and Scattering of Sagittarius A* at 86 GHz: First VLBI with ALMA}
\shorttitle{The Size, Shape, and Scattering of Sagittarius A*}
\author{S.~Issaoun}
\affil{\radboud} \affil{\cfa}
\author{M.~D.~Johnson}
\affil{\cfa}
\author{L.~Blackburn}
\affil{\cfa}
\author{C.~D.~Brinkerink}
\affil{\radboud}
\author{M.~Mo{\'s}cibrodzka}
\affil{\radboud}
\author{A.~Chael}
\affil{\cfa}
\author{C.~Goddi}
\affil{\radboud}\affil{\leiden}
\author{I.~Mart{\'i}-Vidal}
\affil{\chalmers}
\author{J.~Wagner}
\affil{\mpifr}
\author{ S.~S.~Doeleman}
\affil{\cfa}
\author{H.~Falcke}
\affil{\radboud}
\author{T.~P.~Krichbaum}
\affil{\mpifr}
\author{K.~Akiyama}
\affil{\nrao}\affil{\haystack}\affil{\naoj}
\author{U.~Bach}
\affil{\mpifr}
\author{K.~L.~Bouman}
\affil{\cfa} 
\author{G.~C.~Bower}
\affil{\asiaa}
\author{A.~Broderick}
\affil{\perimeter}
\author{I.~Cho}
\affil{\kasi}\affil{\ust}
\author{G.~Crew}
\affil{\haystack}
\author{J.~Dexter}
\affil{\mpe}
\author{V.~Fish}
\affil{\haystack}
\author{R.~Gold}
\affil{\frankfurt}\affil{\perimeter}
\author{J.~L.~G{\'o}mez}
\affil{\granada}
\author{K.~Hada}
\affil{\naoj}
\author{A.~Hern{\'a}ndez-G{\'o}mez}
\affil{\unam}\affil{\irap}
%\author{M.~Hecht} ask him?
%\affil{\haystack} 
\author{M.~Jan{\ss}en}
\affil{\radboud}
\author{M.~Kino}
\affil{\naoj}
\author{M.~Kramer}
\affil{\mpifr}
\author{L.~Loinard}
\affil{\unam}\affil{\unamm}
\author{R.-S.~Lu}
\affil{\shanghai}\affil{\mpifr}
\author{S.~Markoff}
\affil{\api}
\author{D.~P.~Marrone}
\affil{\arizona}
\author{L.~D.~Matthews}
\affil{\haystack}
\author{J.~M.~Moran}
\affil{\cfa}
\author{C.~M{\"u}ller}
\affil{\radboud}\affil{\mpifr}
\author{F.~Roelofs}
\affil{\radboud}
\author{E.~Ros}
\affil{\mpifr}
\author{H.~Rottmann}
\affil{\mpifr}
\author{S.~Sanchez}
\affil{\iram}
\author{R.~P.~J.~Tilanus}
\affil{\radboud}\affil{\leiden}
\author{P.~de Vicente}
\affil{\yebes}
\author{M.~Wielgus}
\affil{\cfa}
\author{J.~A.~Zensus}
\affil{\mpifr}
\author{G.-Y.~Zhao}
\affil{\kasi}

\received{October 19, 2018}

\accepted{December 7, 2018}

\begin{abstract}
The Galactic Center supermassive black hole Sagittarius A* (\sgra) is one of the most promising targets to study the dynamics of black hole accretion and outflow via direct imaging with very long baseline interferometry (VLBI). At 3.5\,mm (86\,GHz), the emission from \sgra\
is resolvable with the Global Millimeter VLBI Array (GMVA). We present the first observations of \sgra\ with the phased Atacama Large Millimeter/submillimeter Array (ALMA) joining the GMVA. Our observations achieve an angular resolution of $\sim$87$\mu$as, improving upon previous experiments by a factor of two.  
We reconstruct a first image of the unscattered source structure of \sgra at 3.5\,mm, mitigating effects of interstellar scattering. The unscattered source has a major axis size of $120 \pm 34\mu$as ($12 \pm 3.4$ Schwarzschild radii), and a symmetrical morphology (axial ratio of $1.2^{+0.3}_{-0.2}$), which is further supported by closure phases consistent with zero within $3\sigma$. We show that multiple disk-dominated models of \sgra\ match our observational constraints, while the two jet-dominated models considered are constrained to small viewing angles. Our long-baseline detections to ALMA also provide new constraints on the scattering of \sgra, and we show that refractive scattering effects are likely to be weak for images of \sgra\ at 1.3\,mm with the Event Horizon Telescope. Our results provide the most stringent constraints to date for the intrinsic morphology and refractive scattering of \sgra, demonstrating the exceptional contribution of ALMA to millimeter VLBI.
\end{abstract}


\section{Introduction}
Supermassive black holes (SMBHs) play a crucial role in shaping our Universe: they evolve symbiotically with their host galaxies and are the cause of extreme environmental changes via accretion, outflows, jets and mergers \citep[e.g.,][]{Ferrarese_2000,Gebhardt_2000}. They are believed to be the origin of the most energetically efficient and powerful processes in the Universe, and yet we are far from fully grasping how these processes are launched and maintained \citep[e.g.,][]{Boccardi_2017, Padovani_2017}. Several theories have been put forward to explain accretion and jet launching mechanisms of SMBHs, but observational evidence to discriminate among theoretical models remains scarce \citep[e.g.,][]{Yuan_Narayan_2014,Fragile_2014}. 

Sagittarius~A* (\sgra) is the radio source associated with the closest known SMBH, with a mass $M \sim 4.1 \times 10^6 M_\odot$, located at the center of our Milky Way, at a distance $D \sim 8.1$ kpc \citep{Ghez_2008,Reid_2009,Gillessen_2009,Gravity_2018}. The angular size of the Schwarzschild radius for \sgra\ is thus estimated to be $R_\mathrm{Sch} = 2GM/c^2 \sim 10\mu$as. Due to its proximity, \sgra\ subtends the largest angle on the sky among all known SMBHs, and is thus the ideal laboratory to study accretion and outflow physics \citep{Goddi_2016}. 

Theoretical models of the dominating component of the radio emission in \sgra\ fall into two broad classes: a relativistic compact jet model or a radiatively inefficient accretion flow \citep{Narayan_1995,Falcke_Markoff_2000,Ozel_2000,Yuan_2003}. However, the southern declination and strong interstellar scattering of \sgra\, (see more details in Section~\ref{sec:scatt}) lead to uncertainty in its intrinsic radio structure, despite decades of centimeter wavelength very long baseline interferometry (VLBI) observations \citep[e.g.,][]{Alberdi_1993,Marcaide_1999,Bower_2004,Shen_2005,Lu_2011a,Bower_2014b}. Consequently, 
these observations have so far been unable to decisively constrain the dominating emission model for \sgra\ to either of those two classes. 
Additional lines of evidence provide support for both models. For instance, frequency-dependent time lags in light-curves of \sgra\ suggest expanding outflows during flares \citep[e.g.,][]{Yusef_2006,Yusef_2008,Brinkerink_2015}. 
VLBI observations at 7\,mm have found evidence for significant intrinsic anisotropy in some epochs \citep{Bower_2014b}, although the anisotropy is not universally seen for other instruments and epochs \citep[e.g.,][]{Zhao_2017}, so the anisotropy may be episodic or may be due to limitations in the scattering mitigation or model fitting procedure. 

VLBI in the mm-regime can reach the smallest spatial scales in \sgra, enabling detection and imaging of the intrinsic structure. At a wavelength of 1.3\,mm, observations with the Event Horizon Telescope (EHT) have shown that the radio emission occurs on scales comparable to the event horizon \citep{Doeleman_2008,Fish_2011,Johnson_2015,Fish_2016,Lu_2018}. On these scales, general relativistic effects such as the ``shadow'' cast by the black hole are expected to determine the source morphology \citep{Falcke_2000}, limiting the view of the innermost accretion flow. At longer wavelengths, scatter-broadening by the interstellar medium (ISM) strongly hinders any attempt to probe intrinsic structure. Observations at 3.5\,mm, where accretion flow kinematics may give rise to an outflow or compact jet and where scatter-broadening becomes subdominant to intrinsic structure, can distinguish between the two classes of models via detailed comparisons of observations and simulations, and help understand the fundamental nature of the radio emission from \sgra. 

The first 3.5\,mm VLBI detection of \sgra, by \cite{Rogers_1994}, gave an initial estimate of the scattered source size using a circular Gaussian fit. \cite{Krichbaum_1998} used three stations to measure the first closure phases (consistent with zero) at 3.5\,mm on a small triangle. Closure phases are a robust observable, since the closed sum of phases in a triangle removes any station-based instrumental effect. A zero value indicates symmetry in the spatial scales probed by the three baselines involved in the closure measurement, a non-zero value implies asymmetry \citep[e.g.,][]{Rauch_2016,TMS}. Subsequent observations, with improved sensitivity and baseline coverage, used closure amplitudes for elliptical Gaussian model-fitting, but the minor axis of the scattered source, along the north-south direction, remained difficult to constrain because of predominantly east-west array configurations  \citep{Doeleman_2001,Shen_2005,Bower_2006,Lu_2011a}. 

The addition of the Large Millimeter Telescope Alfonso Serrano (LMT) and the Robert C. Byrd Green Bank Telescope (GBT) enabled more precise estimates of the intrinsic size and shape of \sgra\ and revealed non-zero closure phases, indicating either intrinsic source asymmetry or substructure from interstellar scattering \citep[][hereafter O16, B16]{Ortiz_2016,Brinkerink_2016}. 
Further analysis by \citet[][hereafter B18]{Brinkerink_2018} found a slight excess of flux density ($\sim$1\% of total flux density) east of the phase center, giving clear deviation from the purely Gaussian geometry that was assumed in model-fitting. Thus, these improved observations support moving beyond simple Gaussian model-fitting to test more complex source models. Imaging is a natural next step, as it does not assume a particular morphological model.

The development of phased-array capability at the Atacama Large Millimeter/submillimeter Array (ALMA) gives unprecedented sensitivity at 3.5\,mm \citep{Doeleman_2010,Fish_2013,Matthews_2018}. In addition to its sensitivity, the geographical location of ALMA provides long north-south baselines to Northern hemisphere sites, probing regions where scattering is sub-dominant to intrinsic structure. In this paper, we present the first VLBI observations of \sgra\ with phased ALMA joining twelve stations of the Global Millimeter VLBI Array (GMVA). 
These observations improve north-south resolution by more than a factor of three compared to previous 3.5\,mm experiments, and they allow us to reconstruct the first unscattered image of \sgra\ at 3.5\,mm. 

The organization of the paper is as follows. In Section~\ref{sec:background}, we give an overview of the relevant background for models of the intrinsic structure and scattering of \sgra. After summarizing the observations and data reduction (Section~\ref{sec:data}) and the imaging (Section~\ref{sec:imaging}), we present our GMVA+ALMA image and discuss data- and image-derived properties of the intrinsic source in the context of previous 3.5\,mm experiments in Section~\ref{sec:results}. In Section~\ref{sec:discussion}, we discuss our new constraints on theoretical models for \sgra\ and its scattering. We summarize our results in Section~\ref{sec:summary}.

\section{Background}\label{sec:background}

\subsection{Theoretical models for Sgr~A* emission}\label{sec:models}

\sgra\ is a bright radio source, with a spectrum that rises with frequency until it peaks near $1\,{\rm mm}$ \citep[e.g.,][]{Falcke_1998,Bower_2015_spectrum}. The long-standing debate on whether the radio/mm emission from \sgra\ is produced by a radiatively inefficient accretion disk or by a relativistic, compact jet present near the black hole \citep[e.g.,][and references therein]{Narayan_1995,Markoff_2007, Moscibrodzka_2014,Ressler_2015,Connors_2017,Davelaar_2018,Chael_2018b} has not been resolved.

Radiative models of \sgra\ based on three-dimensional general relativistic magnetohydrodynamic (GRMHD) simulations of Kerr black hole accretion naturally combine the disk and jet scenarios. Electrons and ions are not in thermal equilibrium in the hot, diffuse \sgra\ accretion flow, therefore simulations with the same gas dynamics (determined by the ions) can have quite different appearances at 3.5\,mm depending on electron thermodynamics assumptions. In particular, both the disk and jet emission dominated models can be realized within a single simulation by adopting a specific distribution for electron heating/acceleration in magnetized plasma in post-processing \citep[e.g.,][]{Moscibrodzka_2013}. Alternatively, electron-ion thermodynamics with a specified prescription for the particle  heating from dissipation can be incorporated self-consistently with the other variables in a single simulation. In this framework, \citet{Ressler_2017} and \citet{Chael_2018b} have shown that both jet- and disk-dominated images can be produced at 3.5\,mm, depending on the underlying physical model for electron heating evolved in the simulation.

These models are mainly used to predict 1.3\,mm EHT observations \citep[e.g.,][]{Chan_2015}. At 1.3\,mm we expect the emission to originate near the event horizon where effects such as gravitational lensing and relativistic Doppler boosting distort any emission into a ring, crescent or a spot-like shape, making any distinction between dominating emission models difficult. At 3.5\,mm, we can potentially constrain the geometry and electron micro-physics of the GRMHD simulations by modeling emission maps in which the physics of accretion rather than relativistic effects shapes the source geometry.

\subsection{Interstellar Scattering of \sgra}\label{sec:scatt}

The index of refraction of a plasma depends on density, so density inhomogeneities in the ionized ISM lead to multi-path propagation of radio waves. The scattering is chromatic, with scattering angles proportional to the squared wavelength of a propagating wave. Because the scattering arises from density irregularities, scattering properties are stochastic by nature; their statistical properties depend on the power spectrum $Q(\mathbf{q})$ of density variations, where $\mathbf{q}$ denotes a wavevector. Along many lines of sight, the scattering is well characterized using a simplified description in which the scattering material is confined within a single thin screen along the line of sight. For background and reviews on interstellar scattering, see \citet{Rickett_1990}, \citet{Narayan_1992}, or \citet{TMS}.

The line of sight to \sgra\ is particularly heavily scattered, as is evidenced by an image with a Gaussian shape and a size that is proportional to wavelength squared for wavelengths $\lambda \gsim 1\,{\rm cm}$ \citep{Davies_1976,vanLangevelde_1992,Bower_2004,Shen_2005,Bower_2006,Johnson_2018}. In addition, the scattering of \sgra\ is anisotropic, with stronger angular broadening along the east-west axis than along the north-south axis \citep{Frail_1994}. The angular broadening has a full width at half maximum (FWHM) of $(1.380 \pm 0.013)\lambda_{\rm cm}^2\,{\rm mas}$ along the major axis and $(0.703 \pm 0.013)\lambda_{\rm cm}^2\,{\rm mas}$ along the minor axis, with the major axis at a position angle $81.9^\circ \pm 0.2^\circ$ east of north \citep[][hereafter J18]{Johnson_2018}. For comparison, the intrinsic source has an angular size of ${\sim}0.4 \lambda_{\rm cm}\, {\rm mas}$ \citepalias{Johnson_2018}, so the ratio of intrinsic size to scatter broadening is ${{\sim}0.3}/\lambda_{\rm cm}$ along the major axis and ${{\sim}0.6}/\lambda_{\rm cm}$ along the minor axis. Consequently, observations at 3.5\,mm are the longest wavelengths with active VLBI for which the intrinsic structure is not sub-dominant to scattering (VLBI observations of \sgra\ at wavelengths between 3.5\,mm and 7\,mm are very difficult because of atmospheric oxygen absorption). 

As discussed by \citet{Psaltis_2018} and \citetalias{Johnson_2018}, the $\lambda^2$ and Gaussian scattering behavior of \sgra\ are universally expected if 1) the intrinsic source size $\theta_{\rm src}$ is subdominant to the scatter broadening angle $\theta_{\rm scatt}$, and 2) the diffractive scale of the scattering $r_{\rm diff} \sim \lambda/\theta_{\rm scatt}$ is smaller than the dissipation scale of turbulence in the scattering material. Thus, even though the angular broadening size and shape are measured very precisely for \sgra\ at centimeter wavelengths, the constraints on the overall scattering properties are quite weak. The expected dissipation scale in the ISM is $10^2 - 10^3\,{\rm km}$ \citep[e.g.,][]{Spangler_Gwinn_1990}, so the expected transition to non-$\lambda^2$ and non-Gaussian scattering (i.e., when the dissipation scale is comparable to the diffractive scale) for \sgra\ occurs at wavelengths of a few millimeters. 
%The limited scattering constraints are especially problematic at millimeter wavelengths because the expected transition to non-$\lambda^2$ and non-Gaussian scattering occurs at a few millimeters. 
Consequently, the scattering properties of \sgra\ measured at centimeter wavelengths cannot be confidently extrapolated to millimeter wavelengths. The uncertainties can be parameterized using physical models for the scattering material, which typically invoke an anisotropic power-law for the power spectrum of phase fluctuations, with the power-law extending between a maximum scale (the outer scale $r_{\rm out}$) and a minimum scale (the inner scale $r_{\rm in}$). In such a generalization, the scattering properties depend on a spectral index $\alpha$, and on the inner scale of the turbulence, $r_{\rm in}$. In this paper, we use the scattering model presented in \citet{Psaltis_2018} with parameters for \sgra\ determined by \citetalias{Johnson_2018}.

The discovery by \citet{Gwinn_2014} of scattering-induced substructure in images of \sgra\ at 1.3\,cm gives an additional constraint on the scattering properties of \sgra. This substructure is caused by modes in the scattering material on scales comparable to the image extent (much larger than $r_{\rm diff}$), so scattering models with identical scatter-broadening may still exhibit strong differences in their scattering substructure. The substructure manifests in the visibility domain as ``refractive noise'', which is an additive complex noise component with broad correlation structure across baselines and time \citep{Johnson_Narayan_2016}. Using observations of \sgra\ from 1.3\,mm to 30\,cm, \citetalias{Johnson_2018} have shown that the combined image broadening and substructure strongly constrains the power spectrum of density fluctuations. However, a degeneracy between $\alpha$ and $r_{\rm in}$ persists, and extrapolating the strength of refractive effects to millimeter wavelengths is still quite uncertain.

Two scattering models effectively bracket the range of possibilities for \sgra. One model (hereafter J18) has a power-law spectral index $\alpha=1.38$ (near the expected value for 3D Kolmogorov turbulence, $\alpha=5/3$) and $r_{\rm in} = 800\,{\rm km}$ (near the expected ion gyroradius in the ionized ISM). The second is motivated by \citet[][herafter GS06]{Goldreich_2006}, who proposed that the scattering of \sgra\ could be caused by thin current sheets in the ISM; it has $\alpha=0$ and $r_{\rm in} \sim 2 \times 10^6\,{\rm km}$. The inner scale in this latter model is several orders of magnitude larger than originally proposed by \citetalias{Goldreich_2006}, but this larger value is required to produce the refractive noise observed at 1.3 and 3.5\,cm. Both the J18 and GS06 models are consistent with all existing measurements of the angular broadening of \sgra\ and with the refractive noise at centimeter wavelengths, but the GS06 model would produce more refractive noise than the J18 model on long baselines at 3.5\,mm, with even more pronounced enhancement for EHT observations \citep[by roughly an order of magnitude; see][]{Zhu_2018}. While long-baseline measurements at 3.5\,mm can discriminate between these possibilities, observations to-date have been inadequate for an unambiguous detection of refractive substructure at this wavelength \citepalias{Ortiz_2016,Brinkerink_2016,Brinkerink_2018}. New observations with ALMA joining 3.5\,mm VLBI, with unprecedented resolution and sensitivity, give the opportunity for long-baseline detections of refractive noise at millimeter wavelengths that can enable discrimination between the two scattering models.

\section{Observations and data reduction}\label{sec:data}
Observations of \sgra\ ($\alpha_\mathrm{J2000} = 17^\mathrm{h}45^\mathrm{m}40^\mathrm{s}.0361$, $\delta_\mathrm{J2000} = -29^\mathrm{\circ}00\mathrm{'}28\mathrm{''}.168$) were made with the GMVA, composed of the eight Very Long Baseline Array (VLBA) antennas equipped with 86\,GHz receivers, the  Green Bank Telescope (GB), the Yebes~40-m telescope (YS), the IRAM 30-m telescope (PV), the Effelsberg 100-m telescope (EB), and the ALMA phased array (AA) consisting of 37 phased antennas. The observations were conducted on 3~April~2017 as part of the first offered VLBI session with ALMA (project code MB007). We recorded a total bandwidth of 256\,MHz per polarization divided in 4 intermediate frequencies (IFs) of 116~channels each. The 12~h track (4~h with the European sub-array and 8~h with ALMA) included three calibrator sources: 1749+096, NRAO~530, and J1924$-$2914. The total integration time on \sgra\ with ALMA was 5.76~h.

\begin{figure}[h!t]
\includegraphics[width=\linewidth]{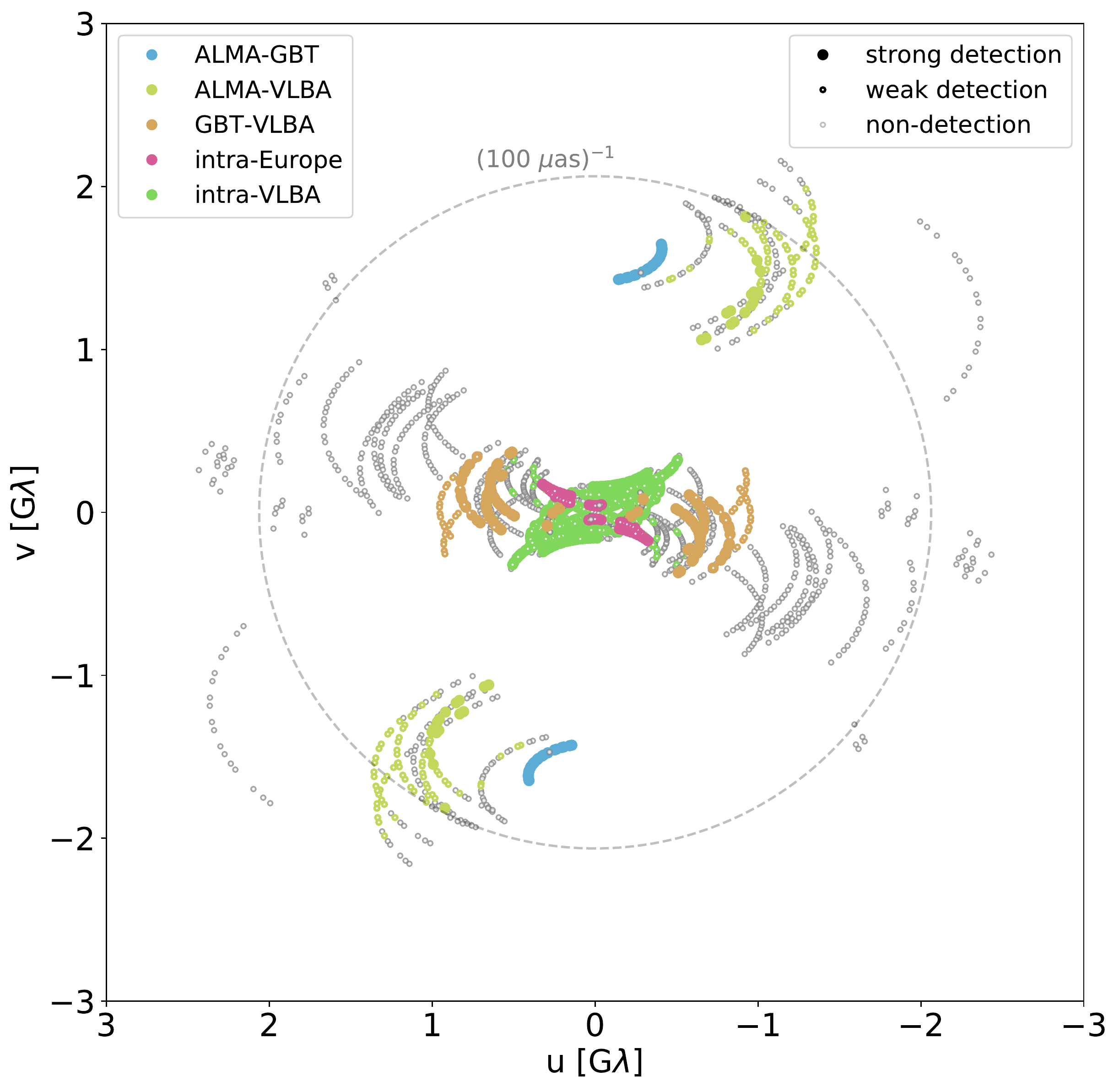}
\includegraphics[width=\linewidth]{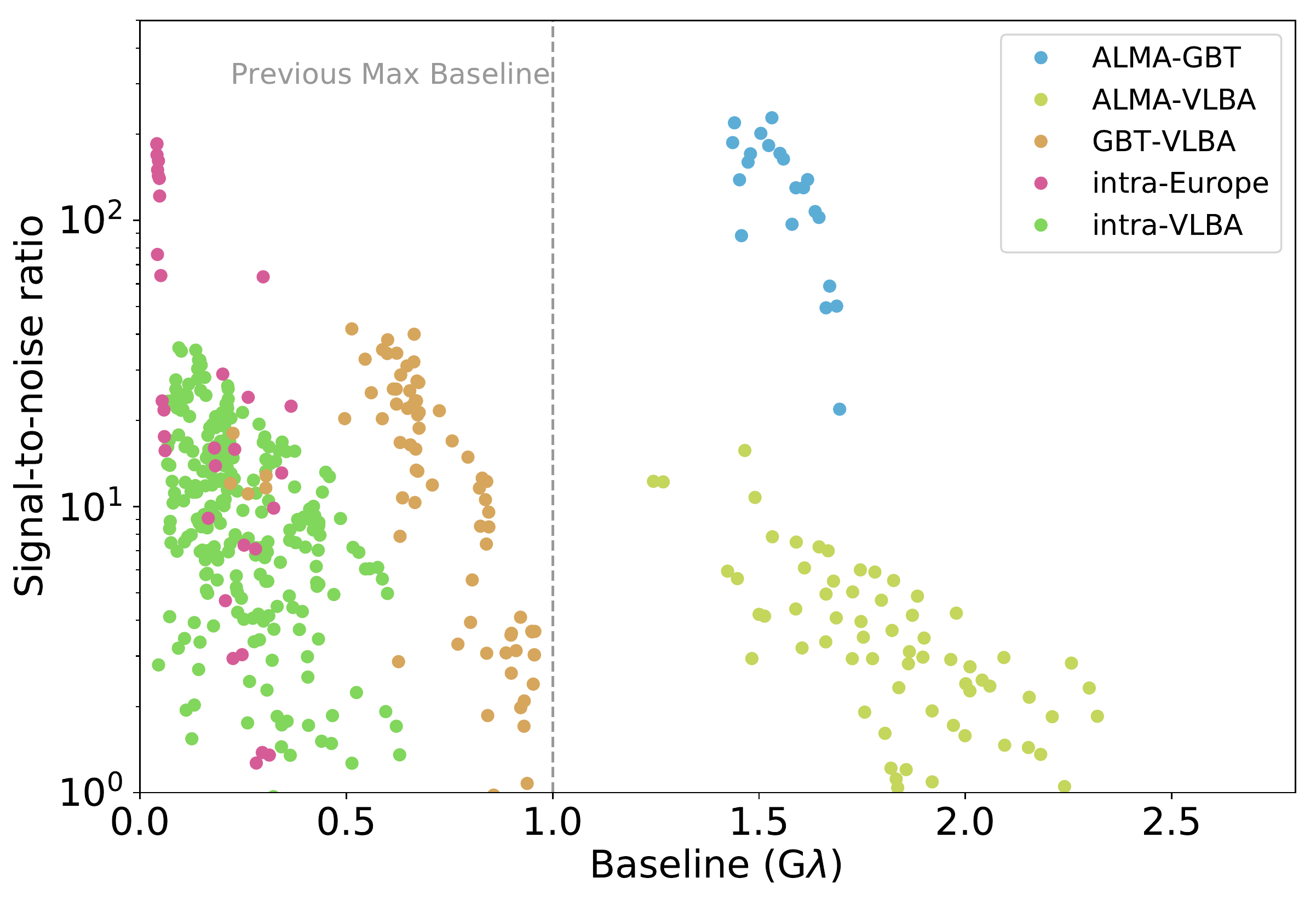}
\caption{{\it Top:} The {\it (u,v)}-coverage of \sgra. Each symbol denotes a scan-averaged measurement: filled colored circles are strong detections; hollow colored circles are weak detections (constrained fringe delay and rate but signal-to-noise ratio (SNR) $<$ 6); and hollow gray circles are non-detections (unconstrained fringe delay and rate) after processing through {\tt HOPS}. {\it Bottom:} The SNR for scan-averaged visibilities on \sgra\ as a function of projected baseline length, showing only detections. All detections beyond ${\sim}1$\,G$\lambda$ are on baselines to ALMA.}
\label{fig:sgra_cov}
\end{figure}

The data were processed with the VLBI correlator at the Max Planck Institute for Radio Astronomy using DiFX \citep{Deller_2011}. After correlation, reduction was carried out using the Haystack Observatory Postprocessing System\footnote{\url{https://www.haystack.mit.edu/tech/vlbi/hops.html}} ({\tt HOPS}) supported by a suite of auxiliary calibration scripts presented in \citet{lindyhops}, with additional validation and cross-checks from the NRAO Astronomical Image Processing System \citep[{\tt AIPS};][]{Greisen_2003}. The {\tt HOPS} software package in its current form arose out of the development of the Mark IV VLBI Correlator, see \citet{Whitney_2004}. During the {\tt HOPS} reduction, ALMA baselines were used to estimate stable instrumental phase bandpass and delay between right and left circular polarization relative to the other stations. ALMA or GBT baselines (depending on signal-to-noise) were used to remove stochastic differential atmospheric phase within a scan. Because atmospheric phase corrections are required on short ($\sim$second) timescales, leading to a large number of free parameters to fit, a round-robin calibration was used to avoid self-tuning: baseline visibility phases on each 58\,MHz IF were estimated using only the remaining 3 IFs, which have independent thermal noise. The integration time for rapid phase corrections was automatically chosen by balancing errors from random thermal variation to those due to atmospheric phase drift, and thus varied with the available signal-to-noise. The median effective integration time was 4.5 seconds. During a final stage of reduction with the {\tt HOPS} fringe fitter {\tt fourfit}, fringe solutions for each scan were fixed to a single set of station-based delays and rates. These were derived from a least-squares solution to baseline detections where unconstrained stations were removed from the dataset. No interpolation of these fringe solutions was performed across scans as the solutions were not stable within their necessary tolerance to maintain coherence. After these phase corrections, our data have enough phase coherence to allow longer averaging times. 

\begin{figure*}[h!t]
\centering
\includegraphics[width=0.41\textwidth]{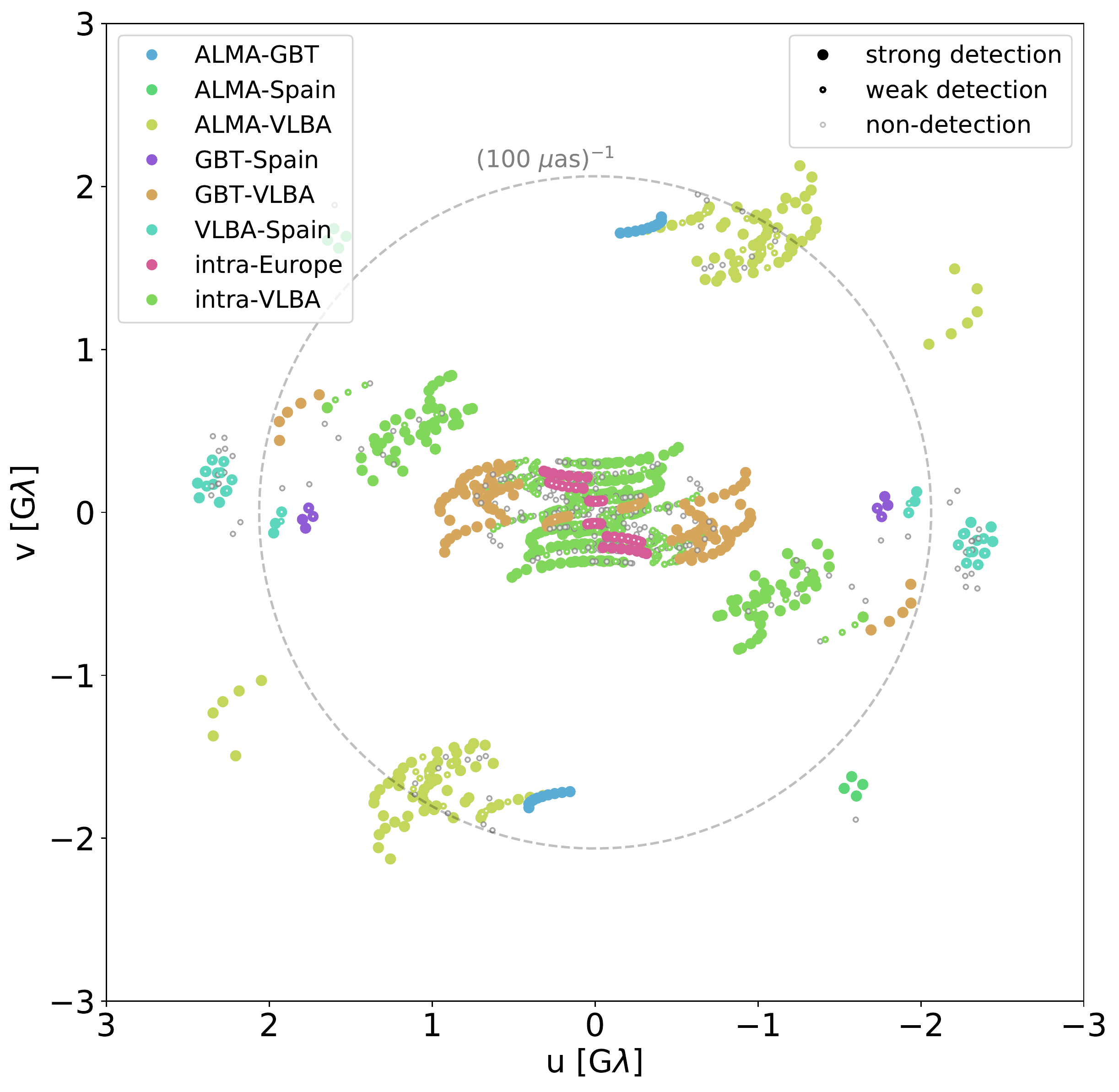}\hspace{0.2cm}
\includegraphics[width=0.52\textwidth]{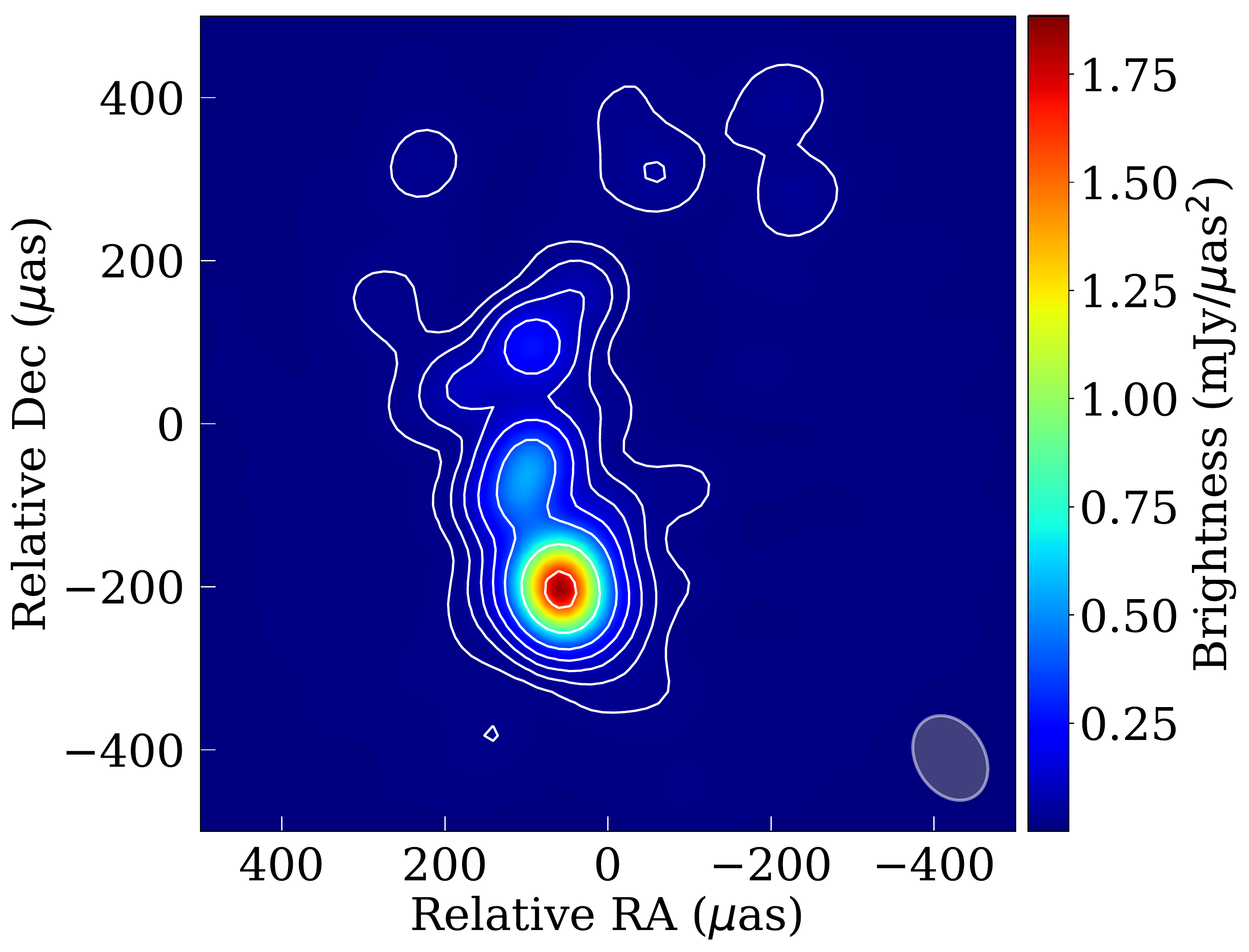}
\caption{ {\it Left}: the ({\it u,v})-coverage of NRAO~530 (symbols are as defined in Figure~\ref{fig:sgra_cov}). {\it Right}: closure-only image of NRAO~530 using the {\tt eht-imaging} library \citep{Chael_2018}, the contour levels start from 1.2\% of the peak and increase in factors of two. The observations have a uniform-weighted beam $ = (111 \times 83)$ $\mu$as,  PA = 32$^\circ$.}
\label{fig:nrao}
\end{figure*}

\begin{figure*}[h!t]
\centering
\includegraphics[width=0.41\textwidth]{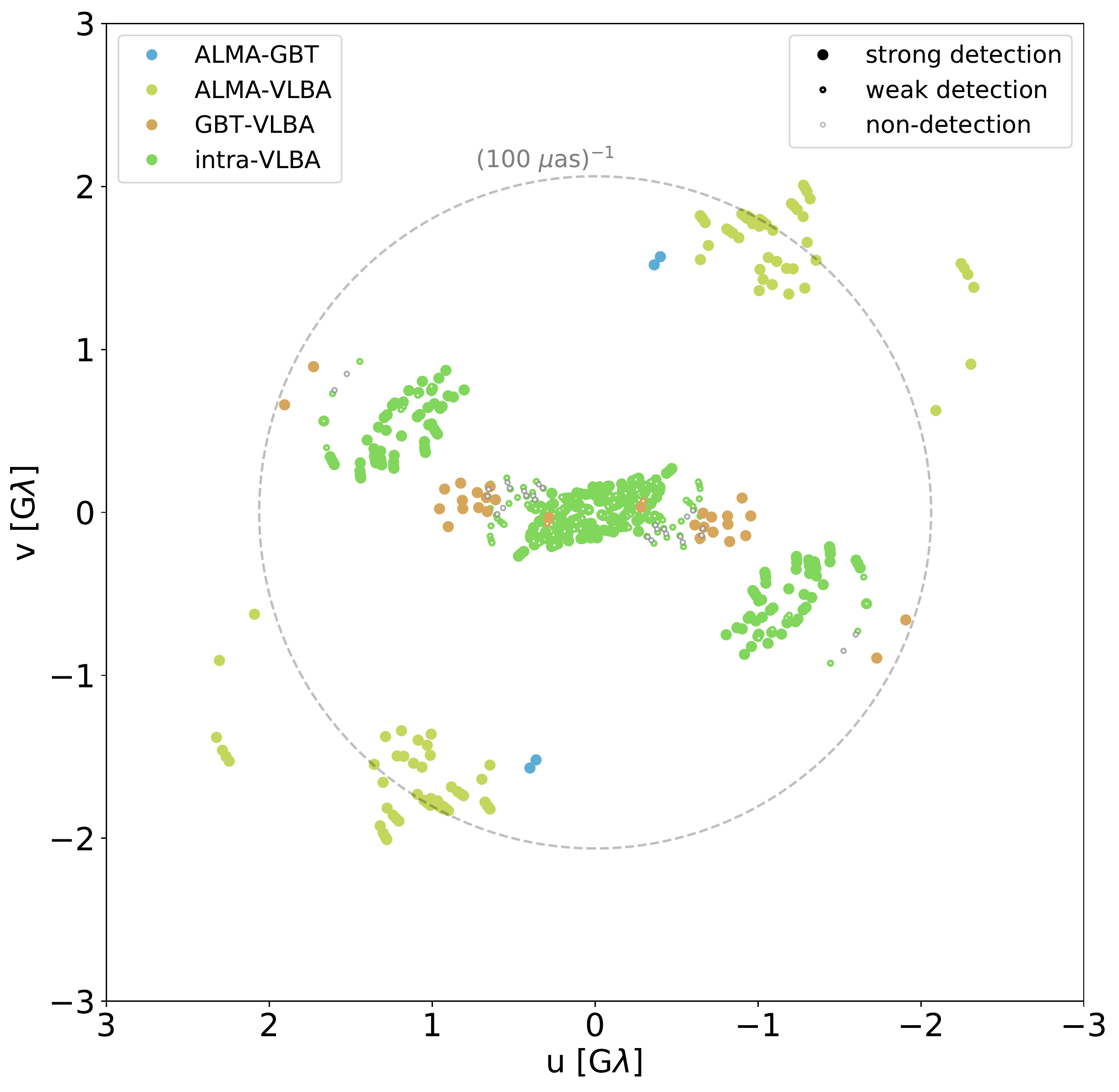}\hspace{0.2cm}
\includegraphics[width=0.52\textwidth]{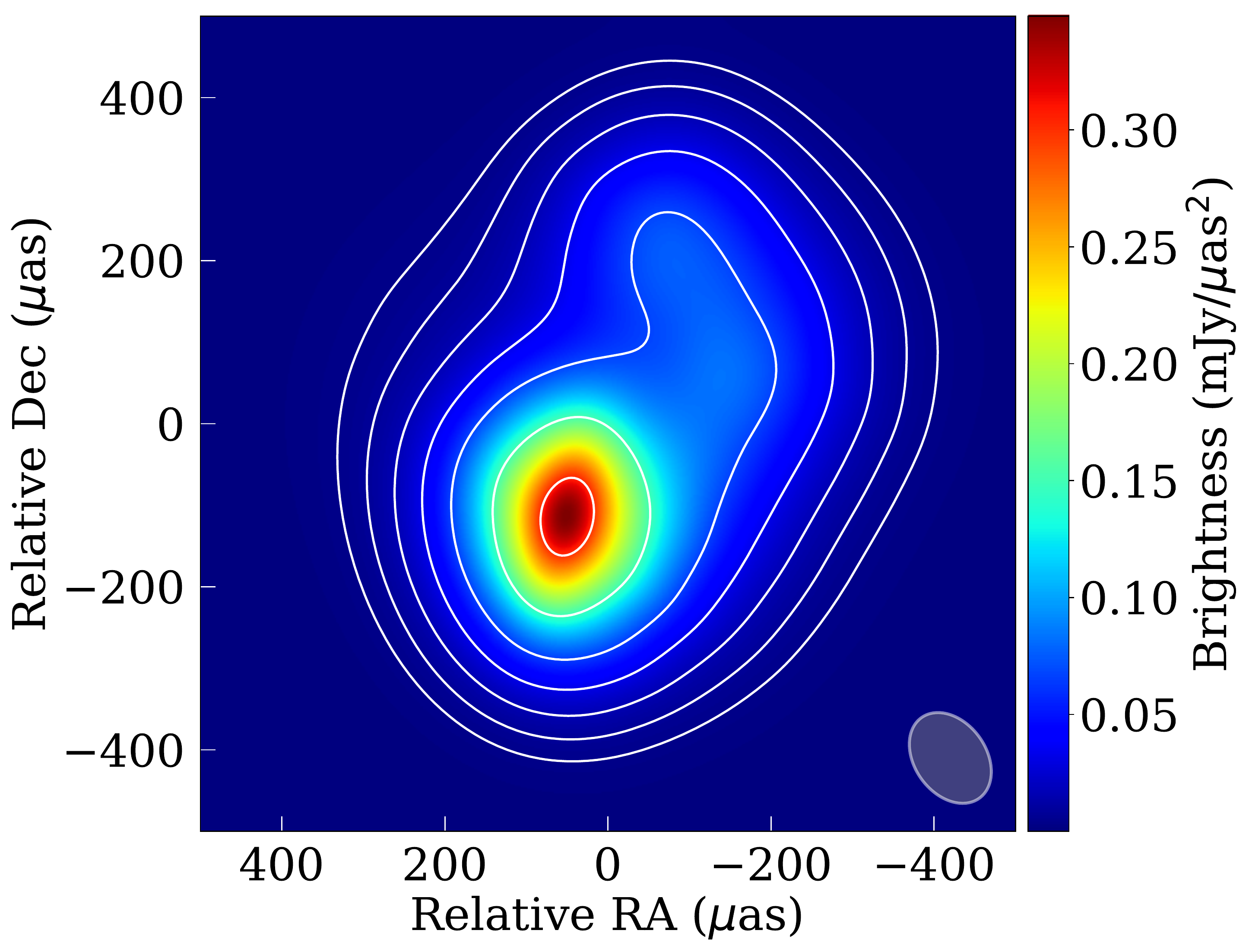}
\caption{ {\it Left}: the ({\it u,v})-coverage of J1924$-$2914 (symbols are as defined in Figure~\ref{fig:sgra_cov}). {\it Right}: closure-only image of J1924$-$2914 using the {\tt eht-imaging} library \citep{Chael_2018}, the contour levels start from 1.2\% of the peak and increase in factors of two. The observations have a uniform-weighted beam $= (122 \times 88)$ $\mu$as, PA = 36$^\circ$. The European stations did not observe this source.}
\label{fig:J1924}
\end{figure*}

We performed a-priori amplitude calibration using provided telescope gain information and measured system temperatures during the observations. The heterogeneity of the stations in the GMVA required us to adopt a careful approach to the amplitude calibration. The calibration for ALMA was fully provided by the ALMA quality assurance (QA2) team \citep{Goddi_2018}, and system equivalent flux densities (SEFDs) were generated with a high time cadence by \texttt{PolConvert} \citep{Marti_2016}. Both YS and PV measure effective system temperatures via the chopper wheel method, and thus do not require an additional opacity correction to their SEFDs. However, the rest of the array (VLBA, GB, EB) measures system temperatures via the noise diode method, requiring an additional opacity correction to account for atmospheric attenuation of the visibility amplitudes. Unfortunately, several VLBA stations observed in difficult weather conditions (ice, wind, rain), leading to limited detections on baselines to Owens Valley (OV), North Liberty (NL) and Pie Town (PT) stations. Additionally, observations at PV suffered from phase coherence losses in the signal chain during the observations, leading to poor quality data and lower visibility amplitudes on those baselines, which cannot be rescaled with a-priori calibration information. Figure~\ref{fig:sgra_cov} shows the detections and non-detections for \sgra\ (top panel) and corresponding signal-to-noise ratio of scan-averaged visibilities for \sgra\ detections. All detections beyond ${\sim}1$\,G$\lambda$ are on baselines to ALMA. After a-priori calibration, we can proceed with imaging routines to determine the morphology of the calibrators and the target source.

\section{Imaging}\label{sec:imaging}
We employ the {\tt eht-imaging} library\footnote{\url{https://
github.com/achael/eht-imaging}}, a regularized maximum likelihood imaging software package, to image our sources \citep{Chael_2016,Chael_2018}. Due to the elevated noise level for the VLBA in our observations and the scattering properties of Sagittarius A*, standard imaging software packages like {\tt AIPS} \citep{Greisen_2003} or {\tt Difmap} \citep{Shepherd_1995} do not offer the flexibility and necessary tools to obtain an unscattered image of the source . The {\tt eht-imaging} library is a {\tt Python}-based software package that is easily scriptable, flexible and modular. It is able to make images with various data products (closure phase and amplitude, bispectra, visibilities), and it contains a suite of image ``regularizers'' such as maximum entropy and sparsity regularization. The library also possesses a routine for ``stochastic optics'', a regularized implementation of scattering mitigation presented in \citet{Johnson_2016}, making it a natural choice for our analysis. In this section we present our imaging methods for both calibrators (Section~\ref{sec:cal_im}) and for \sgra\ (Section~\ref{sec:sgra_im}).

\subsection{Calibrators NRAO~530 and J1924$-$2914}\label{sec:cal_im}

Both NRAO~530 and J1924$-$2914 appear point-like to ALMA when acting as a connected-element interferometer ($\sim${70}~k$\lambda$, $\sim${3}~arcseconds resolution), with NRAO~530 having a flux density of $2.8 \pm 0.3$~Jy and J1924$-$2914 having a flux density of $5.0 \pm 0.5$~Jy (as measured by interferometric-ALMA). Even on the angular scales probed by VLBI, both sources are very compact and stable, making them ideal for imaging. The operational difficulties and poor weather conditions at the VLBA were largely offset by the high sensitivity of ALMA. The extent of all detections is shown in the left panel of Figure~\ref{fig:nrao} for NRAO~530 and Figure~\ref{fig:J1924} for J1924$-$2914. A third calibrator was also observed, 1749+096, but only for a few minutes with the full array, and is thus omitted from further analysis.

The large number of detections on both NRAO~530 and J1924$-$2914 led to a correspondingly large number of closure phases and closure amplitudes. We thus imaged both sources using only closure quantities, following the method from \citet{Chael_2018}, constraining the total flux of the image to match measurements from interferometric-ALMA. We present images of the two calibrators in Figure~\ref{fig:nrao} and Figure~\ref{fig:J1924} (right panels). The morphology of NRAO 530 is consistent with previous observations of the source \citep{Bower_1997,Bower_1998,Feng_2006,Chen_2010, Lu_2011b}. The elongation of the J1924$-$2914 jet in the north-west direction at 86 GHz is consistent with mm-jet studies from previous observations at 43 GHz by \citet{Shen_2002} and 230 GHz by \citet{Lu_2012}. These two sources are common calibrators for \sgra. They are therefore particularly useful to study at multiple frequencies to adequately calibrate observations at 1.3\,mm from the EHT. 

\begin{table}[h!t]
\caption{Station median multiplicative gains to the visibility amplitudes.}
\label{tab:gains}
\begin{center}
\begin{tabular}{cccc}
\hline 
\hline 
{\bf Station} & {\bf \sgra} & {\bf NRAO~530} & {\bf J1924$-$2914} \\
\hline
{\bf BR} & 2.2$^{+1.5}_{-0.8}$  & 1.7$^{+0.5}_{-0.5}$  & 2.0$^{+1.4}_{-1.8}$  \\
{\bf FD} & 2.2$^{+1.2}_{-0.6}$  & 1.9$^{+1.2}_{-0.5}$  & 1.8$^{+0.7}_{-0.4}$  \\
{\bf GB} & 1.2$^{+1.7}_{-0.4}$  & 1.1$^{+0.5}_{-0.1}$  & 1.2$^{+0.7}_{-0.4}$  \\
{\bf KP} & 2.4$^{+2.2}_{-0.6}$  & 2.2$^{+1.2}_{-0.4}$  & 2.1$^{+2.4}_{-0.4}$  \\
{\bf LA} & 2.2$^{+2.8}_{-1.0}$  & 1.9$^{+1.7}_{-0.7}$  & 2.9$^{+2.0}_{-2.1}$  \\
{\bf NL} & 4.6$^{+13.3}_{-2.1}$  & 4.7$^{+9.7}_{-1.5}$  & 5.0$^{+22.6}_{-2.4}$  \\
{\bf OV} & 1.9$^{+3.1}_{-1.0}$  & 1.9$^{+0.9}_{-0.6}$  & 1.7$^{+1.6}_{-0.3}$  \\
{\bf PT} & 11.4$^{+2.2}_{-5.3}$  & 19.3$^{+17.8}_{-13.3}$  & 12.9$^{+36.7}_{-8.4}$  \\
\hline
\end{tabular}
\end{center}

NOTE--- Median (and 95th percentile) multiplicative gains to the visibility amplitudes for common stations from the two calibration methods: 1) self-calibration of \sgra\ amplitudes below {0.75}G$\lambda$ to the Gaussian source estimated from \citetalias{Ortiz_2016,Brinkerink_2018}, and 2) self-calibration of NRAO~530 and J1924$-$2914 observations to the images produced with closure phases and closure amplitudes. The European stations and ALMA are not shown as they are not self-calibrated for all three sources. We flagged NL and PT due to their high median gain and erratic gain solutions.
\end{table}

\begin{figure*}[h!t]
\centering 
\includegraphics[width=0.7\linewidth]{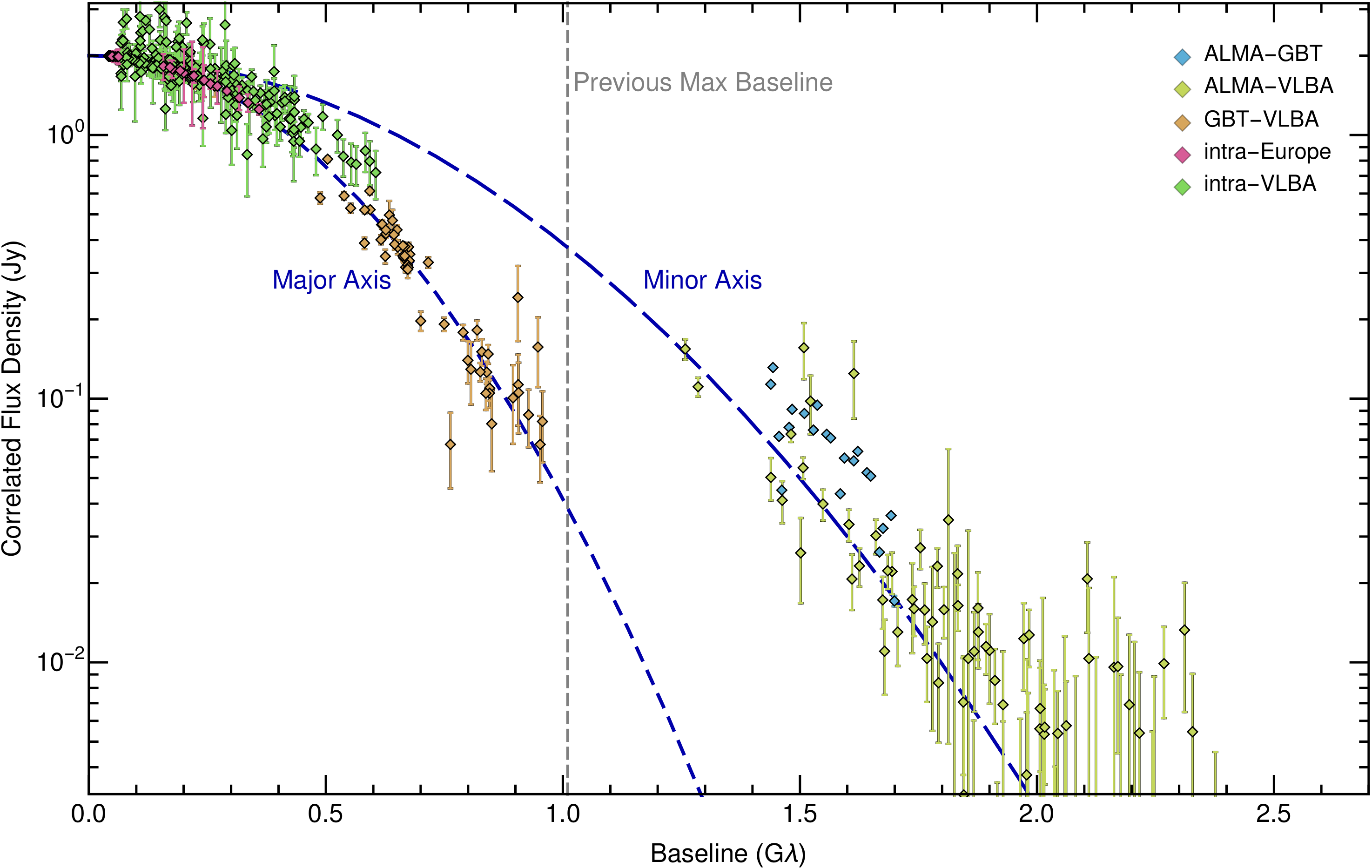}
\caption{Noise-debiased correlated flux density of \sgra\ as a function of projected baseline length for data after self-calibrating to the Gaussian source from \citetalias{Ortiz_2016} and \citetalias{Brinkerink_2018} using only baselines shorter than $0.75\,{\rm G}\lambda$. Because the a-priori calibration for the GBT was excellent (see Table~\ref{tab:gains}), we did not apply the derived GBT gains. Dashed dark blue curves show expected visibilities along the major and minor axes for an anisotropic Gaussian source with FWHM of 215~$\mu$as by 140~$\mu$as (the source size from \citetalias{Ortiz_2016} and \citetalias{Brinkerink_2018}). All detections beyond ${\sim}1$G$\lambda$ are baselines to ALMA, and all show marked deviations from the Gaussian curves.}
\label{fig:sgra_vis}
\end{figure*}

\subsection{Sagittarius A*}\label{sec:sgra_im}
\subsubsection{Self-calibration}

We obtained far fewer detections on \sgra\ than on the calibrators, and our detections also had lower signal-to-noise ratio (SNR). Consequently, we did not have enough information to synthesize images of \sgra\ using only closure quantities. Moreover, due to the suboptimal performance of the VLBA (bad weather, signal loss likely from pointing issues), additional amplitude calibration was necessary to mitigate severe signal losses at various stations. 

We utilized two methods for amplitude calibration: 
\begin{enumerate}
    \item we self-calibrated to closure-only images of NRAO 530 and J1924$-$2914 to obtain smoothed station gain trends,
    \item we self-calibrated all \sgra\ visibility amplitudes within 0.75G$\lambda$ (predominantly intra-VLBA measurements) using an anisotropic Gaussian visibility function determined by previous 3.5\,mm experiments \citepalias{Ortiz_2016,Brinkerink_2018}, with the total flux set by the interferometric-ALMA measurement.
\end{enumerate}

For the second method, we used a visibility function corresponding to a Gaussian source size of 215 by {140}\,$\mu$as with a position angle of {80}$^\circ$ (east of north) and a total flux density of 2.0$ \pm 0.2$\,{Jy}. The choice of the Gaussian size is motivated by similar results obtained for \citetalias{Ortiz_2016} and \citetalias{Brinkerink_2018} taken one month apart, showing stable source dimensions. Both these experiments had the high sensitivity of the LMT, adding north-south coverage to recover the minor axis size with greater accuracy than older experiments. In our interferometric-ALMA measurements, \sgra\ has flux density variations at the 10\% level on a timescale of about 4 hours, not significantly affecting our static imaging. Note that gains were derived by self-calibration using only short baselines, but because they are station-based, they were then applied to correct visibility amplitudes on longer baselines as well. 

The two methods gave comparable gain solutions, hence validating the Gaussian assumption for short-baseline measurements (Table~\ref{tab:gains} shows median multiplicative station gains to the visibility amplitudes). We flagged the VLBA stations NL and PT, which showed extreme signal loss in both methods. The GBT performed well for all three sources, so we chose to keep the original a-priori calibration. Because GBT is only linked to NL in the inner 0.75 G$\lambda$ baseline cut for \sgra, the derived gains for GBT introduce large variations to the ALMA-GBT amplitudes that come from difficulty locking NL gains due to its bad weather. Ignoring the self-calibration solutions gave more stable amplitudes on the ALMA-GBT baseline.

Figure~\ref{fig:sgra_vis} shows the scan-averaged visibilities for \sgra\ after self-calibration of the inner 0.75~G$\lambda$ baselines to the Gaussian source size (method 2). All detections above 1~G$\lambda$ are new measurements to ALMA. The ALMA-GBT baseline has significantly higher flux density than expected from the minor axis of the previously fitted Gaussian source size from \citetalias{Ortiz_2016} and \citetalias{Brinkerink_2018}. VLBA detections to ALMA show clear deviations from Gaussian behavior.

\begin{figure*}[h!t]
\includegraphics[width=0.5\textwidth]{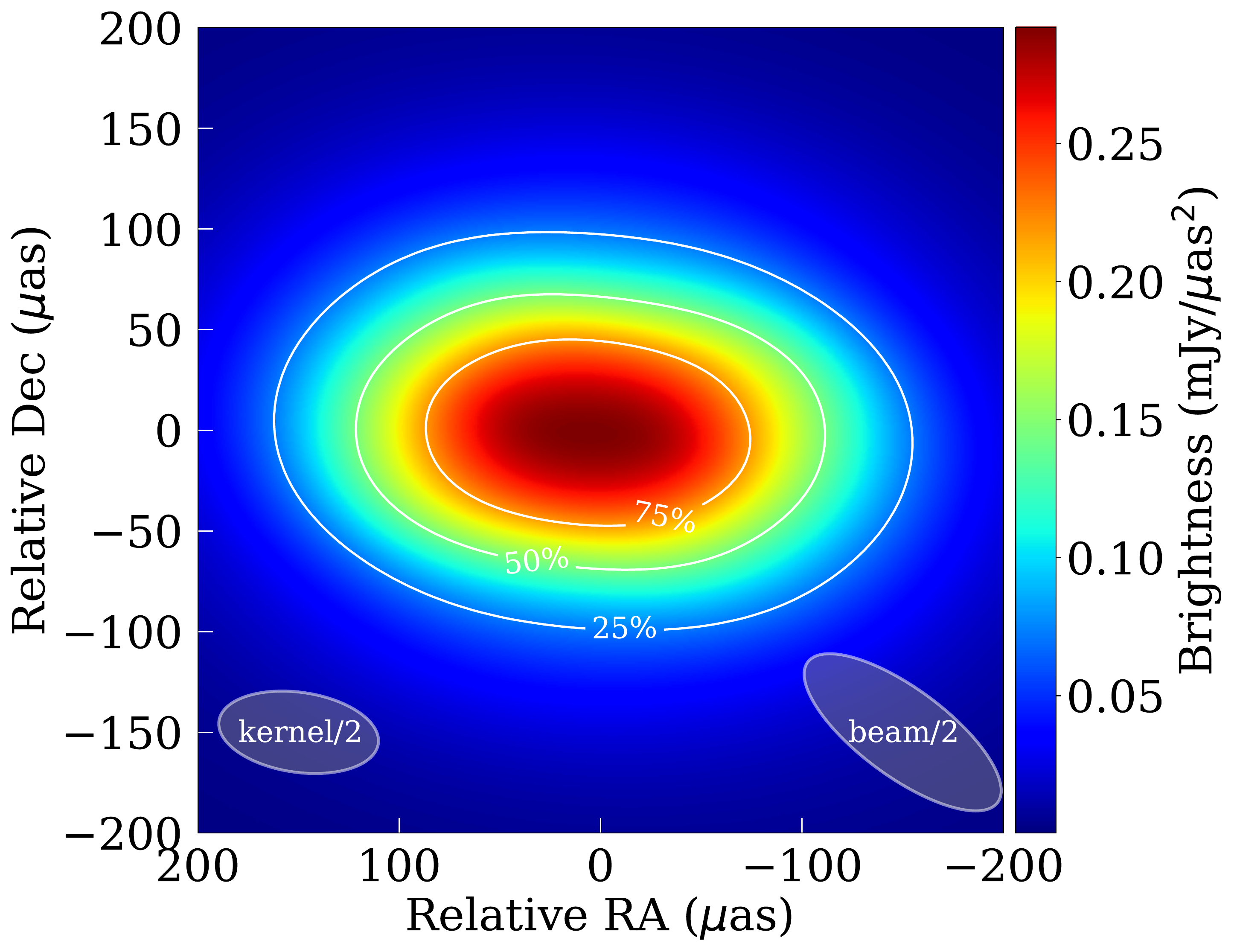}\hspace{0.2cm}
\includegraphics[width=0.49\textwidth]{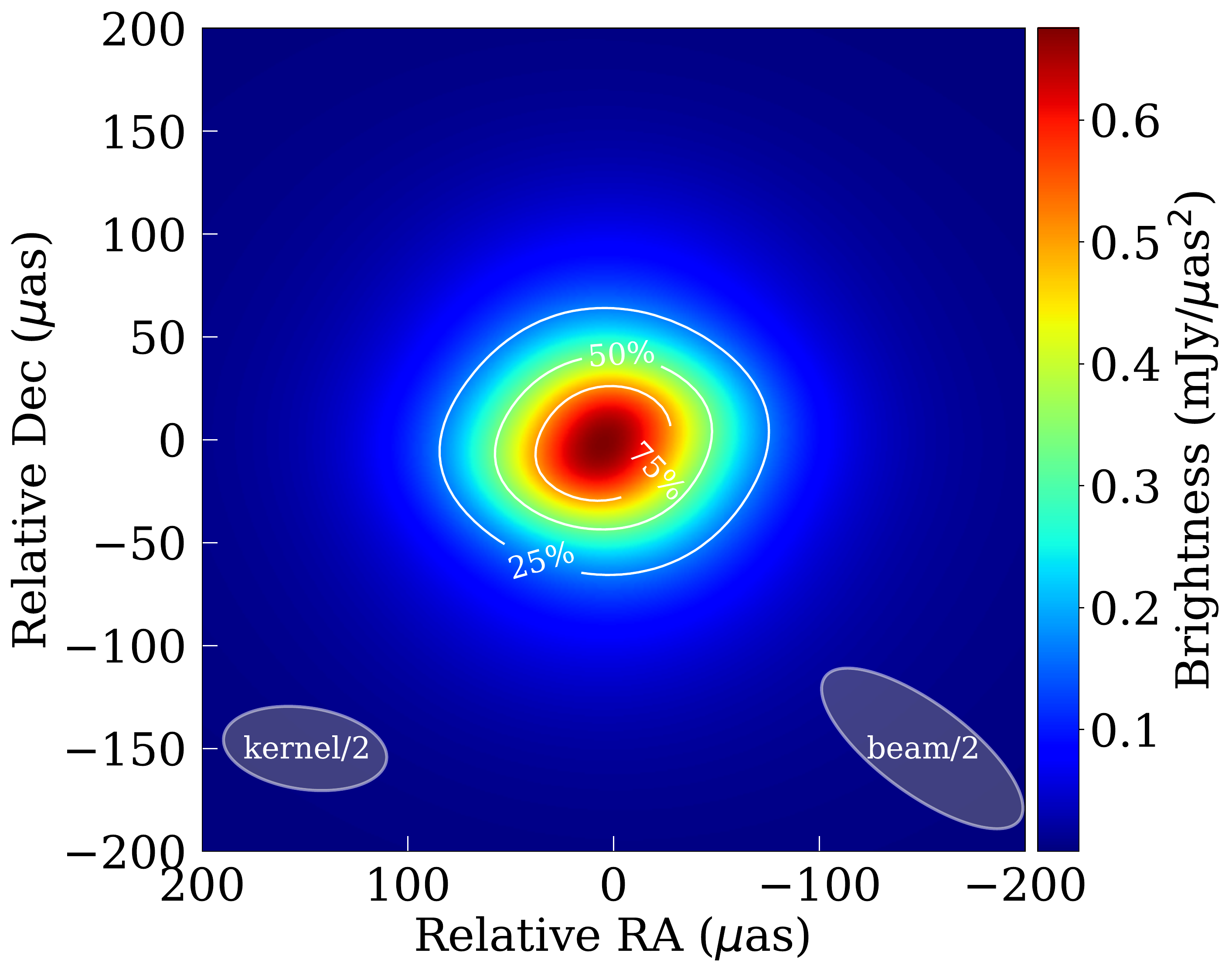}
\caption{{\it Left}: the scattered image of \sgra, reconstructed with the second moment regularizer and stochastic optics ($\theta_{\mathrm{maj}}=228 \pm 46~\mu$as, $\theta_{\mathrm{min}}=143 \pm 20~\mu$as from LSQ). {\it Right}: the reconstructed image from stochastic optics \citep{Johnson_2016} of the intrinsic source ($\theta_{\mathrm{maj}}=120 \pm 34~\mu$as, $\theta_{\mathrm{min}}=100 \pm 18~\mu$as from LSQ). In each panel, the ellipses at the bottom indicate half the size of the scatter-broadening kernel ($\theta_{\mathrm{maj}}=159.9~\mu$as, $\theta_{\mathrm{min}}=79.5~\mu$as, PA = 81.9$^\circ$) and of the observing beam.}
\label{fig:sgra_so}
\end{figure*}

\begin{figure*}[h!t]
\begin{minipage}[b][\textheight]{\linewidth}
\centering 
\includegraphics[width=\linewidth]{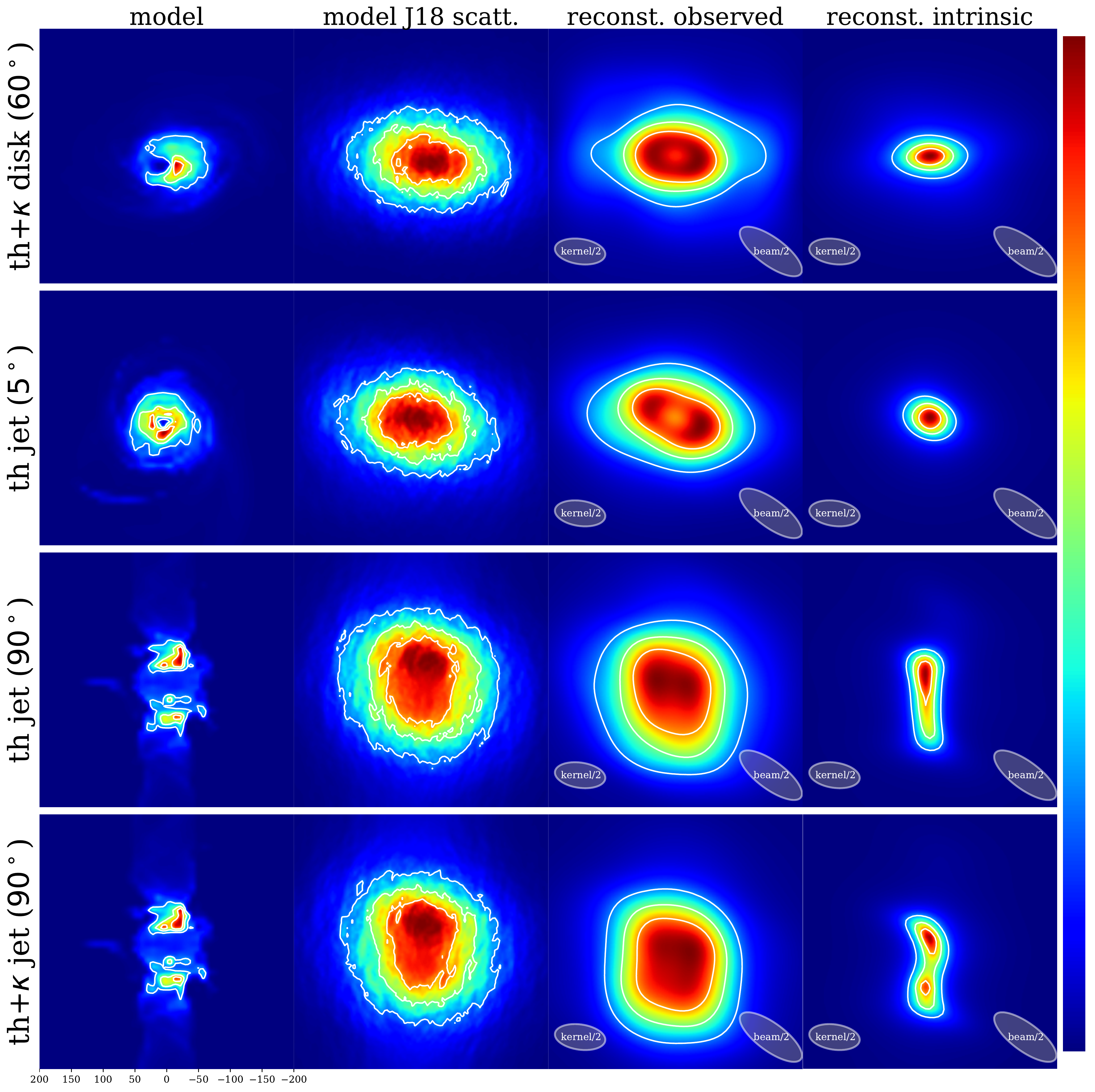}
\caption{Model and reconstructed images from four example 3D GRMHD models, plotted here in linear scale. The contour levels represent 25, 50 and 75\% of the peak flux. The first column shows the original model images as given from simulations: ``th+$\kappa$ disk'' is a thermal disk model with 1\% accelerated particles in a power-law ($\kappa$) distribution; ``th jet'' is a thermal jet model; ``th+$\kappa$ jet'' is a thermal jet model with 10\% accelerated particles in a $\kappa$ distribution \citep{Moscibrodzka_2009,Moscibrodzka_2014,Moscibrodzka_2016,Davelaar_2018}. The inclinations of the models are given in the parentheses.  The second column shows the model images scattered with the J18 scattering model: these are the images sampled to make the simulated observations. The third column shows the observed (scattered) image reconstructed with the second moment regularizer and stochastic optics, and the fourth column shows the reconstructed image from stochastic optics of the corresponding intrinsic source. In the third and fourth columns, the ellipses at the bottom indicate half the size of the scatter-broadening kernel and of the observing beam.}
\label{fig:model_images}
\end{minipage}
\end{figure*}
%\newpage 
\subsubsection{Imaging with regularized maximum likelihood}

The performance of the VLBA impaired our ability to model-fit to the dataset and obtain an accurate source size estimate using only short baselines (i.e., baselines that do not heavily resolve the source). In addition, large measurement uncertainties for the visibility amplitudes on intra-VLBA baselines made image convergence difficult and unstable. We thus implemented a new imaging regularization: we constrained the second central moment of the image to match more robust measurements of the scattered source size from Gaussian model fitting to previous observations \citepalias{Ortiz_2016,Brinkerink_2018}. If we think of the centroid (first moment) of the image as the mean position of the emission, its variance (or second moment) is the spread of emission from the mean, equivalent to the extent of the source along its principal axes~\citep{Hu_1962}. The regularization is equivalent to constraining the curvature of the visibility function at zero baseline. This method helps to calibrate short-baseline visibilities during the imaging process, while allowing long-baseline detections to ALMA to still recover smaller scale structure in our images. 
This method is now included and implemented in the {\tt eht-imaging library} via gradient descent minimization (the effects and fidelity of the regularizer will be presented in \citet{Issaoun_2018}). We also made use of the ``stochastic optics'' scattering mitigation code from \cite{Johnson_2016} to disentangle the effects of scattering and produce the intrinsic image of \sgra.

To reach our final result, we first imaged the scattered source using closure quantities and visibility amplitudes (with equal weights).
The regularizers used in the scattered image, with a weighting of 10$\%$ of the data weights, were: Gull-Skilling maximum entropy; total squared variation; and second moment regularization with the second moment matrix given by that of the Gaussian used for self-calibration. 
Each of these regularizers favors particular image features, while enforcing image positivity and a total flux density constraint. 
Gull-Skilling entropy favors pixel-by-pixel similarity to the prior image (we used the previously fitted Gaussian source as the prior). Total-squared variation regularization favors small image gradients, producing smooth edges (see \citet{Chael_2018} for a detailed discussion of these regularizers). Second moment regularization constrains the second derivative of the visibility function at the zero baseline (which is proportional to the second central image moment) to match a specified value; we thereby constrained our short baselines to match those of the Gaussian source measured in previous experiments \citepalias{Ortiz_2016,Brinkerink_2018} without imposing assumptions on the visibilities measured by longer baselines, which reflect image substructure.  In the scattering mitigation code, the second moment regularization is only applied to the observed image, such that the intrinsic image derived by the scattering deconvolution is not directly constrained by the regularizer but still remains within physical size ranges. After imaging with closure quantities and corrected visibility amplitudes, we then self-calibrated the visibility phases and amplitudes to the obtained scattered image before imaging with stochastic optics (using the same regularization parameters).

The stochastic optics framework is implemented within the {\tt eht-imaging} library via regularized maximum likelihood. The code solves for the unscattered image by identifying, separating and mitigating the two main components of the scattering screen, introduced in Section~\ref{sec:background}: small-scale diffractive modes that blur the image, causing the ensemble-average scattered image to be a convolution of the true image and the scattering kernel (predominantly east-west scatter-broadening); and large-scale refractive modes that introduce stochastic image substructure (ripples distorting the image). The code simultaneously solves for the unscattered image and the large-scale phase screen causing refractive scattering, while assuming a given model for the diffractive blurring kernel and the refractive power spectrum $Q({\mathbf q})$ (governing the time-averaged scattering properties). In our case, we used the scattering kernel (with a size of (159.9 $\times$ 79.5)\,$\mu$as, PA of 81.9$^\circ$) and power spectrum (with $\alpha=1.38$ and $r_\mathrm{in}=800$\,km) from the J18 scattering model. See \citet{Johnson_2016} for a more detailed description of the method.
Two iterations of stochastic imaging and self-calibration are done for convergence. We present in Figure~\ref{fig:sgra_so} our resulting intrinsic and scattered images of \sgra.

\begin{deluxetable*}{ccccccc}[!ht]
\tablecaption{Comparison of the true size and the derived size from imaging from synthetic datasets for four simulated images.  \label{tab:image_test}}
\tablehead{ \colhead{\bf Model} & \colhead{Method} & \colhead{ }  & \colhead{$\theta_\mathrm{maj}$ ($\mu$as)} & \colhead{$\theta_\mathrm{min}$ ($\mu$as)}& \colhead{Axial ratio} & \colhead{PA (deg)}
 }
\startdata 
\multirow{6}{*}{\bf Th+$\kappa$ disk (60$^\circ$)} & \multirow{3}{*}{$2^\mathrm{nd}$ mom.} & True & 121.0 & 97.3 & 1.24 & 105.4 \\
& { }& Image & 184.0 & 131.8 & 1.4 & 87.6 \\
& { }& Difference & 63.0 ($0.4\theta_\mathrm{beam}$)  & 34.5 ($ 0.1 \theta_\mathrm{beam}$) & 0.16 ($ 0.1 \theta_\mathrm{beam}$) & 17.8 ($0.6\theta_\mathrm{beam}$) \\
& \multirow{3}{*}{LSQ} & True & 79.7 & 77.0 & 1.04 & 109.9 \\
& { }& Image & 101.9 & 59.6 & 1.7 & 0.8 \\
& { }& Difference & 22.2 ($ 0.1 \theta_\mathrm{beam}$) & 17.4 ($ 0.1 \theta_\mathrm{beam}$) & 0.66 ($ 0.1 \theta_\mathrm{beam}$) & 69.3 ($\gtrsim 0.4 \theta_\mathrm{beam}$) \\
\hline 
\multirow{6}{*}{\bf Th jet (5$^\circ$)} & \multirow{3}{*}{$2^\mathrm{nd}$ mom.} & True & 112.5 & 99.0 & 1.14 & 13.8 \\
& { }& Image & 148.7 & 124.8 & 1.19 & 74.2 \\
& { }& Difference & 36.2 ($0.3 \theta_\mathrm{beam}$) & 25.8 ($0.1\theta_\mathrm{beam}$) & 0.05 ($0.02\theta_\mathrm{beam}$) & 60.4 ($\gtrsim \theta_\mathrm{beam}$) \\
& \multirow{3}{*}{LSQ} & True & 88.0 & 81.2 & 1.08 & 179.7 \\
& { }& Image & 65.5 & 51.9 & 1.26 & 158.4 \\
& { }& Difference & 22.5 ($0.2\theta_\mathrm{beam}$) & 29.3 ($0.1\theta_\mathrm{beam}$) & 0.18 ($0.03\theta_\mathrm{beam}$) & 21.3 ($0.2\theta_\mathrm{beam}$) \\
\hline 
\multirow{6}{*}{\bf Th jet (90$^\circ$)} & \multirow{3}{*}{$2^\mathrm{nd}$ mom.} & True & 174.0 & 65.8 & 2.64 & 179.8 \\
& { }& Image & 178.1 & 135.3 & 1.32 & 176.4 \\
& { }& Difference & 4.1 ($0.02\theta_\mathrm{beam}$) & 69.5 ($0.5\theta_\mathrm{beam}$) & 1.32 ($0.6\theta_\mathrm{beam}$) & 3.4 ($0.3\theta_\mathrm{beam}$) \\
& \multirow{3}{*}{LSQ} & True & 160.8 & 63.2 & 2.54 & 178.8 \\
& { }& Image & 130.3 & 42.4 & 3.07 & 177.1 \\
& { }& Difference & 30.5 ($0.2\theta_\mathrm{beam}$) & 20.8 ($0.1\theta_\mathrm{beam}$) & 0.53 ($0.04\theta_\mathrm{beam}$) & 1.7 ($0.2\theta_\mathrm{beam}$) \\
\hline 
\multirow{6}{*}{\bf Th+$\kappa$ jet (90$^\circ$)} & \multirow{3}{*}{$2^\mathrm{nd}$ mom.} & True & 182.4 & 65.7 & 2.78 & 179.7 \\
& { }& Image & 177.5 & 127.6 & 1.4 & 177.6 \\
& { }& Difference & 4.9 ($0.02\theta_\mathrm{beam}$) & 61.9 ($0.4\theta_\mathrm{beam}$) & 1.38 ($0.6\theta_\mathrm{beam}$) & 2.1 ($0.2\theta_\mathrm{beam}$) \\
& \multirow{3}{*}{LSQ} & True & 166.6 & 62.9 & 2.65 & 178.7  \\
& { }& Image & 141.5 & 49.9 & 2.83 & 179.2 \\
& { }& Difference & 25.1 ($0.1\theta_\mathrm{beam}$) & 13.0 ($\pm 0.1\theta_\mathrm{beam}$) & 0.18 ($0.02\theta_\mathrm{beam}$) & 0.5 ($0.1\theta_\mathrm{beam}$) \\
\enddata
\tablecomments{In each case, we compute the sizes using two methods: directly from the image second central moment (``$2^\mathrm{nd}$ mom.''), and from a 2D Gaussian fit to the image with least-squares minimization (``LSQ''). We give the absolute difference between the true and estimated values and also express the difference as a fraction of the projected beam FWHM $\theta_\mathrm{beam}$ along the measured axis, or as the fraction of the propagated error from the beam-widths on both axes for the axial ratio. The uncertainty on the position angle (PA) is expressed as the fraction of one-dimensional beam blurring of the image for which the standard deviation in PA with blurring along different directions matches the difference between the true and measured PA (see text for additional details).}
\end{deluxetable*}

\subsubsection{Uncertainties of image-derived parameters}\label{sec:tests}

To determine the uncertainties in the imaging method and size measurements for \sgra, we performed imaging tests on simulated observations where the intrinsic model image was known. 
We tested our imaging method on four snapshots from 3D GRMHD simulations of \sgra at 86\,GHz \citep{Moscibrodzka_2009,Moscibrodzka_2014,Moscibrodzka_2016,Davelaar_2018}, using the same sampling, coverage and noise as our observations. The model images were scattered with the J18 scattering model and sampled with our GMVA+ALMA coverage, before being imaged via the same imaging routine applied to the \sgra\ data described above. 

While the imaging procedure is identical, these reconstructions do have some advantages relative to our reconstruction of the actual observations. For example, we used the ensemble-average properties of the J18 scattering model as inputs to the scattering mitigation: i.e., we assume perfect knowledge of the diffractive scattering kernel and the time-averaged power spectrum. We also measure the second moment of the scattered simulated images and use it as an exact input to the second moment regularization. Because the scattering is subdominant to intrinsic structure and because the second moment is estimated to excellent accuracy in previous experiments, we do not expect either of these effects to significantly advantage the reconstructions of simulated data.

In Figure~\ref{fig:model_images} we present the original 3D GRMHD model images, the model images scattered with the J18 scattering model (as observed in the simulated observations), and the reconstructed observed (scattered) and intrinsic images from the imaging method. In Table~\ref{tab:image_test} we compare the true intrinsic source sizes from the models to the intrinsic source sizes derived from the imaging routine. We determined the source size parameters using two methods: first by measuring the second central moment of the image ($2^\mathrm{nd}$ mom.) and deriving Gaussian parameters; and second by doing a 2D Gaussian fit with a least-squares minimization (LSQ) onto the image. 

\begin{deluxetable*}{lcccccccc}[!ht]
\tablecaption{Observed and intrinsic sizes for \sgra\ at 86 GHz. \label{tab:sgra_size}}

\tablehead{ \colhead{\multirow{2}{*}{\bf Reference}} & \colhead{$\theta_\mathrm{maj,obs}$ } & \colhead{$\theta_\mathrm{min,obs}$ }  & \colhead{PA$_\mathrm{obs}$ } & \colhead{Axial ratio} & \colhead{$\theta_\mathrm{maj,int}$ } & \colhead{$\theta_\mathrm{min,int}$ }&  \colhead{PA$_\mathrm{int}$} & \colhead{Axial ratio} \\
\colhead{ } & \colhead{($\mu$as)} & \colhead{ ($\mu$as)}  & \colhead{(deg)} & \colhead{$r_\mathrm{obs}$} & \colhead{($\mu$as)} & \colhead{ ($\mu$as)}&  \colhead{ (deg)} & \colhead{$r_\mathrm{int}$}
 }
\startdata 
\cite{Rogers_1994} & $150 \pm 50$ & - & - & - & $<130$ & - & - & - \\
\cite{Krichbaum_1998} & $190 \pm 30$ & - & - & - & - & -& - & -\\
\cite{Doeleman_2001} & $180 \pm 20$ & - & - & - & $<130$ & - & - & - \\
\cite{Shen_2005} & $210^{+20}_{-10}$ & $130^{+50}_{-13}$ & $79^{+12}_{-33}$ & $1.6^{+0.4}_{-0.5}$ & - & -& - & -\\
\cite{Lu_2011a} & $210 \pm 10$ & $130 \pm 10$ &  $83 \pm 2$ & $1.6 \pm 0.1$ & $139 \pm 17$& $102 \pm 21$ & - &  $1.4 \pm 0.3$ \\
\citetalias{Ortiz_2016} BD183C & $213 \pm 2$ & $138 \pm 4$ & $81 \pm 2$ & $1.54 \pm 0.04$ & $142 \pm 9$ & $114 \pm 15$ & - & $1.2 \pm 0.2$\\
\citetalias{Ortiz_2016} BD183D & $222 \pm 4$ & $146 \pm 4$ & $75 \pm 3$ & $1.52 \pm 0.05$ & $155 \pm 9$ & $122 \pm 14$ & - & $1.3 \pm 0.2$ \\
\citetalias{Brinkerink_2018} (clos.amp.) & $215.1 \pm 0.4$ & $145 \pm 2$ & $77.9 \pm 0.4$ & $1.48 \pm 0.01$ & - & - & - & -\\
\citetalias{Brinkerink_2018} (selfcal) & $217 \pm 22$ & $165 \pm 17$ & $77 \pm 15$ & $1.3 \pm 0.2$ & - & - & - & -\\
\citetalias{Johnson_2018} BD183C & $215 \pm 4$ & $139 \pm 4$ & $81 \pm 3$ & $1.55 \pm 0.05$ & $143^{+11}_{-12}$ & $114 ^{+7}_{-8}$ & - & $1.25^{+0.20}_{-0.16}$ \\
This work ($2^\mathrm{nd}$ mom.)$^*$ & $239 \pm 57$ & $172 \pm 103$ & $84 \pm 2$ & $1.4^{+1.1}_{-0.4}$ & $176 \pm 57$ & $152 \pm 103$ & $85.2 \pm 44 ^{**}$ &  $1.2^{+1.1}_{-0.2}$ \\
This work (LSQ)$^*$ & $228 \pm 46$& $143 \pm 20$ & $86 \pm 2$ & $1.6 \pm 0.3$ & $120 \pm 34$ & $100 \pm 18$ & $96.0 \pm 32^{**}$ & $1.2^{+0.3}_{-0.2}$ \\
\enddata
\tablecomments{$^*$Image-domain size estimates. The stated uncertainties are derived using the largest parameter errors for reconstructions of simulated images.\\
$^{**}$Position angle estimates are not meaningfully constrained because of the near symmetry of the major/minor axes.}
\end{deluxetable*}

Next, we evaluate the difference between true and reconstructed image parameters. We sought to define an approach that quantifies these differences in a way that is related to the reconstructed image properties and the observing beam. When expressed in this way, we can use parameter errors on these reconstructed simulated images to predict uncertainties on parameters derived from our reconstructed image with data. 

To this end, Table~\ref{tab:image_test} expresses the difference between the true and measured source major and minor axes as a fraction of the projected beam FWHM $\theta_\mathrm{beam}$ along the corresponding axis. For the axial ratio, we express the difference between the true and measured ratios as a fraction of the cumulative error from both axes (the projected beam-widths along the measured major and minor axes added quadratically). 

However, while it is straightforward and well-motivated to express uncertainties on axis lengths and their ratio in terms of the observing beam, uncertainty on the position angle (PA) is more subtle. We opted to create an ensemble of beam-convolved reconstructed images and to use the scatter in the PA of the ensemble as an estimate of the PA uncertainty. The ensemble of images is constructed by convolving the single reconstructed image with an ensemble of narrow beams, sampling all position angles. Each of these beams has a major axis size given by the projected observing beam size along the same position angle and a minor axis size of zero. We thereby stretch the image along each direction, up to the extent of the observing beam, and examine the overall dependence of the reconstructed image on this stretching. With this approach, images that are nearly isotropic will have large PA uncertainty, while highly elongated images (relative to the beam size) will have small PA uncertainty.

In general, we find that the LSQ method fares better than $2^\mathrm{nd}$ moment for determining the source parameters, likely due to weak extended flux in the images skewing the second moment parameters to larger values. As expected, both methods perform poorly when determining the position angle of a fairly symmetrical source, for which it remains largely unconstrained. However, for more elongated source geometry, both methods are able to accurately recover the intrinsic position angle. We adopt the LSQ method to quantify the size of \sgra\, via image-domain fitting. Although the Gaussian approximation does not describe fully our source morphology, it is suitable for comparisons to visibility-domain model fits from previous observations of \sgra\, presented in Section~\ref{sec:results}.

\section{Results}\label{sec:results}
\subsection{Intrinsic source constraints from imaging}
 Figure~\ref{fig:sgra_so} shows the unscattered and scattered images of \sgra, as imaged following the method described in Section~\ref{sec:imaging}. The (uniform-weighted) beam size of the \sgra\ observations is $(235 \times 87)$ $\mu$as, with a position angle (east of north) of 53.6$^\circ$. While the shorter baselines of the array (intra-VLBA, VLBA-GBT, intra-European) see primarily a Gaussian source elongated in the east-west direction, longer baselines are expected to pick up on non-Gaussian source structure or refractive noise from interstellar scattering. In this particular observation, our longest baselines are mainly north-south to ALMA (see Figure~\ref{fig:sgra_cov}), where scattering has less of an effect on the source. As seen in Figure~\ref{fig:sgra_so}, left panel, the reconstructed scattered image looks very smooth and Gaussian-like, showing no obvious refractive noise in the image. We also see a similar outcome in our imaging tests, presented in Section~\ref{sec:tests}. Although the scattered images (second column in Figure~\ref{fig:model_images}) have visible ripples of scattering substructure, the reconstructed scattered images (third column) appear very smooth. This is likely because our GMVA+ALMA observations sample low levels of refractive noise mainly along the north-south direction, whereas our east-west sensitivity and resolution do not provide adequate detections of scattering substructure to be able to reconstruct the fine structure in the scattered images. Thus the low level of refractive noise detected on our ALMA baselines does not produce visible distortions in the reconstructed scattered image.

We present the measured source sizes using our two methods ($2^\mathrm{nd}$ mom. and LSQ) in Table~\ref{tab:sgra_size}, along with historical measurements and estimates. The uncertainties are conservative estimates taken from the largest relative uncertainties on the parameters of simulated images for each method (see Table~\ref{tab:image_test}). We assume a Gaussian source geometry for size estimates and comparisons, but this may not be the correct source model. As seen in the example images (Figure~\ref{fig:model_images}), the true and reconstructed intrinsic images are not Gaussian, therefore this choice of parametrization is only to simplify comparisons with previous measurements and simulations. We find that our source size measurements are consistent with previous observations and indicate the source dimensions and small asymmetry are persistent across multiple years. 

Lastly, we note that uncertainties in the intrinsic size caused by remaining uncertainties in the scattering kernel are quite small ($\lsim 10\,\mu{\rm as}$), even allowing for the full range of uncertainty on $\alpha$ and $r_{\rm in}$ \citepalias{Johnson_2018}. The reason they are small is because the scattering parameters for angular broadening are estimated to an accuracy of a few percent, and because the intrinsic structure is not subdominant to scatter broadening.

\begin{figure}[h!t]
\centering 
\includegraphics[width=0.99\linewidth]{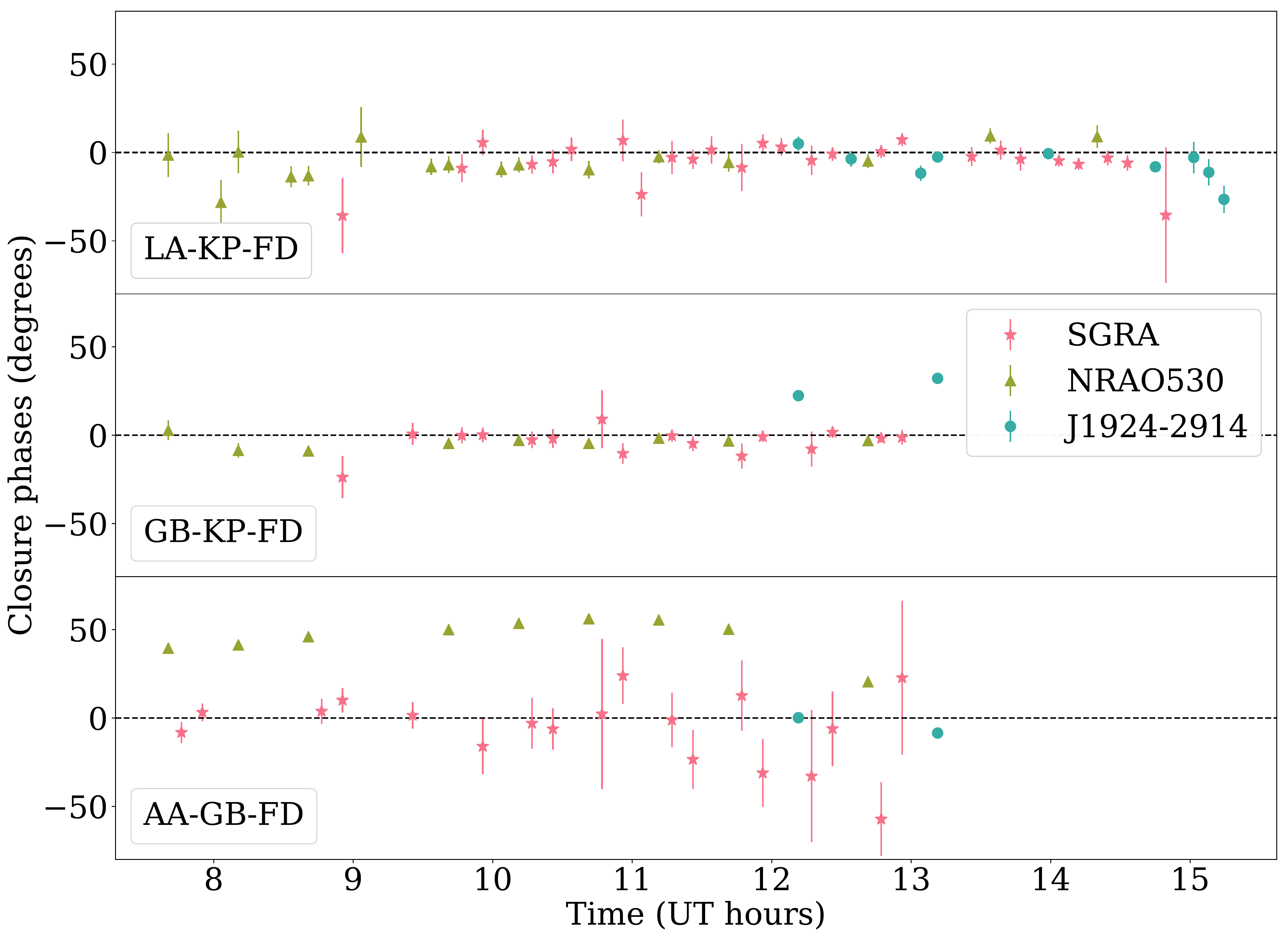}
\caption{Scan-averaged closure phases for \sgra, NRAO~530 and J1924$-$2914 on three triangles (LA-KP-FD, GBT-KP-FD, ALMA-GBT-FD) formed after processing through {\tt HOPS}. The larger uncertainties on the ALMA-GBT-FD triangle are primarily because of low correlated flux density on the ALMA-FD baseline (see Figure~\ref{fig:sgra_vis}). Non-zero closure phase indicates source asymmetry. Although NRAO~530 and J1924$-$2914 show significant deviations from zero, all \sgra\ closure phases are consistent with zero within 3$\sigma$.
}
\label{fig:sgra_cphase}
\end{figure}

\subsection{Intrinsic source constraints from closure phases}
Closure phases provide an alternative and complementary assessment of source asymmetry directly from observations. They are weakly affected by refractive scattering and are unaffected by station-based calibration issues. Thus, they offer robust information on the intrinsic properties of \sgra. 

We computed closure phases for all sources from scan-averaged visibilities. The GMVA+ALMA array contains 13 stations, yielding many triangles with a wide range of sizes. As seen in Figures~\ref{fig:nrao} and \ref{fig:J1924}, there are multiple long-baseline detections to ALMA on calibrators that do not appear for \sgra\ (Figure~\ref{fig:sgra_cov}). We thus selected three example triangles of different sizes and orientations that are present for the two main calibrator sources (NRAO~530 and J1924$-$2914) and with multiple detections for \sgra.

We present in Figure~\ref{fig:sgra_cphase} the closure phases on three representative triangles: a small intra-VLBA (LA-KP-FD) triangle; an east-west medium-sized triangle to GBT (GB-KP-FD); and a long north-south triangle to ALMA (AA-GB-FD). Although all three triangles provide robust detections for all three sources, with non-zero closure phases for the calibrators, \sgra\ closure phases remain very close to zero: the weighted mean closure phase on AA-GB-FD is $-1.1 \pm 2.4^\circ$; the weighted mean closure phase on GB-KP-FD is $-1.7 \pm 1.1^\circ$; and the weighted mean closure phase on LA-KP-FD is $-1.8 \pm 1.1^\circ$. The largest closure phases on all three triangles deviate from zero by less than 3$\sigma$. 

\citetalias{Ortiz_2016} and \citetalias{Brinkerink_2016} detected small non-zero closure phases (${\lsim}10^\circ$) on triangles including the highly sensitive LMT and/or GBT. These non-zero closure phases were observed on triangles not present in our GMVA+ALMA observations, and they probed different scales and directions from our new predominantly north-south triangles with ALMA. Deviations of a few degrees, as observed by \citetalias{Ortiz_2016} and \citetalias{Brinkerink_2016}, fall within our confidence bounds due to low signal-to-noise on VLBA baselines, and thus would not be detectable with our current observations. Moreover, the geometrical models to describe the asymmetry in \citetalias{Brinkerink_2016} produce closure phases on our triangles that would be indistinguishable from zero with our current measurements.  Thus, our results are consistent with previous observations of \sgra. 

\section{Discussion}\label{sec:discussion}

\begin{figure*}[!ht]
\centering
\includegraphics[width=0.8\linewidth]{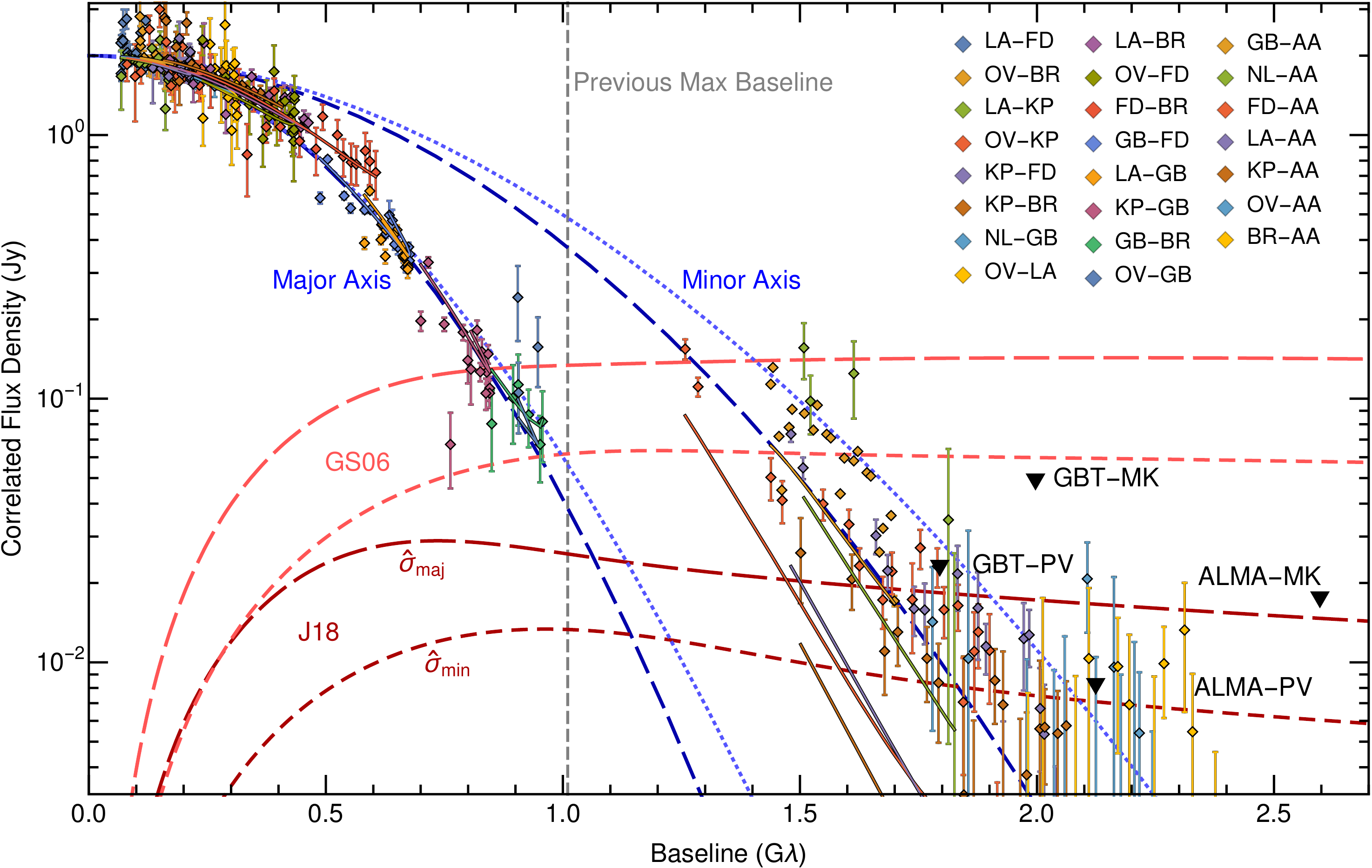}
\caption{Noise-debiased correlated flux density for \sgra\ as a function of projected baseline length for data after self-calibrating to the Gaussian source from \citetalias{Ortiz_2016,Brinkerink_2018} using only baselines shorter than $0.75\,{\rm G}\lambda$. Because the a-priori calibration for the GBT was excellent (see Table~\ref{tab:gains}), we did not apply the derived GBT gains. Baseline labels are ordered by median baseline length. Intra-European baselines are entirely constrained by the self-calibration and are omitted here for clarity. Dark blue dashed curves show expected visibilities along the major and minor axes for an anisotropic Gaussian source with FWHM of $(215\,\mu{\rm as}, 140\,\mu{\rm as})$; light blue dotted curves show the visibility expected for an anisotropic intrinsic Gaussian source $(140\,\mu{\rm as}, 100\,\mu{\rm as})$ scattered with the non-Gaussian kernel from the J18 scattering model, which has an image size (via $2^{\rm nd}$ mom.) of $(216\,\mu{\rm as}, 132\,\mu{\rm as})$; red curves show the expected renormalized refractive noise along the major and minor axes for the J18 and GS06 scattering models. Detections on baselines longer than $1{\rm G}\lambda$ are only obtained for baselines oriented close to the minor axis of the scattering kernel (all are ALMA-VLBA/GBT). Labeled black triangles show upper limits ($4\sigma$) on four sensitive baselines at other orientations, all of which have corresponding detections for our calibrators. Colored lines show the anisotropic Gaussian model curves for the corresponding data.}
\label{fig:Refractive_Noise}
\end{figure*}

\subsection{Constraints on the Refractive Scattering of \sgra}

Our longest baselines heavily resolve the scattered image of \sgra\ while also providing exceptional sensitivity (especially baselines to ALMA). Therefore they are sensitive to a non-Gaussian scattering kernel (from a finite inner scale) and to ``refractive noise,'' which corresponds to image substructure introduced by interstellar scattering. In this section, we use our long-baseline measurements to constrain scattering models for \sgra.

Figure~\ref{fig:Refractive_Noise} shows our detected correlated flux density as a function of baseline length. The ALMA-GBT baseline, probing scales along the minor axis of the source, measures significantly higher correlated flux density than predicted from the Gaussian curves from model fitting to shorter baseline data, shown as the dark blue dashed curves in Figure~\ref{fig:Refractive_Noise} \citepalias{Ortiz_2016,Brinkerink_2018}. This enhancement could either indicate non-Gaussian intrinsic structure (e.g., a compact core with a diffuse halo) or a non-Gaussian scattering kernel (requiring an inner scale $r_{\rm in} \ll 1.5 \times 10^9 \lambda \approx 5000\,{\rm km}$). For example, the ALMA-GBT measurements are comparable to the values predicted for an anisotropic Gaussian intrinsic source combined with the J18 scattering model, which has $r_{\rm in} = 800\,{\rm km}$, shown as the light blue dotted curves in Figure~\ref{fig:Refractive_Noise}. 

We also detect correlated flux density on baselines that are expected to entirely resolve the scattered source. Here, the enhanced flux density indicates the presence of image substructure that can either be intrinsic or scattering-induced. For scattering substructure, the signal is expected to be significantly stronger for baselines that are aligned with the major axis of the scattering (see Figure~\ref{fig:Refractive_Noise}). The two candidate scattering models presented in Section~\ref{sec:scatt} (with different spectral index $\alpha$ and inner scale $r_\mathrm{in}$ governing the refractive noise power spectrum) predict different levels of refractive noise along both the major and minor axes of the scattering: the GS06 model predicts, on average, nearly one order of magnitude more correlated flux density on long baselines than the J18 model. However, our most sensitive detections (ALMA-VLBA/GBT) are along the minor axis of the scattering. 

The mean visibility amplitude (after debiasing to account for thermal noise) on baselines longer than 1.8\,G$\lambda$ is 6\,mJy. Because this amplitude may contain contributions from both scattering substructure and intrinsic substructure, it only determines an upper limit on the level of refractive noise from scattering substructure. Moreover, even if there were no intrinsic substructure contribution on these baselines, the 6\,mJy signal would still not directly determine the level of refractive noise because refractive noise is stochastic; the inner 95\% of visibility amplitudes sampled on a single baseline over different scattering realizations will fall in the range $[0.16, 1.9] \times \hat{\sigma}$, where $\hat{\sigma}$ is the RMS ``renormalized'' refractive noise \citepalias[i.e., refractive noise after removing the contributions of flux modulation and image wander, which our observations would absorb into the overall calibration; see][]{Johnson_2018}. 

We can tighten the constraints on refractive noise by combining samples from many baselines, although these will be correlated \citep[see][]{Johnson_Narayan_2016}. Following the Monte Carlo approach of \citetalias{Johnson_2018}, we find that combining baselines longer than $1.8\,{\rm G}\lambda$ gives a 95\% confidence range for the mean amplitude of refractive noise on a baseline with $(u,v) = (1.167, -1.638) \times 10^9\lambda$ of 3-18\,mJy if the 6\,mJy of correlated flux density is entirely from refractive noise. For comparison, the J18 model predicts a mean refractive noise amplitude of approximately 7\,mJy on this baseline, while the GS06 model predicts a mean refractive noise of 60\,mJy on this baseline. Thus, the GS06 model is incompatible with our measurements. The GS06 model also significantly over-predicts the signal on our baselines oriented closer to the major axis, for which our measurements only provide upper limits (labeled black triangles in Figure~\ref{fig:Refractive_Noise}).

If the minor axis detections are from scattering substructure, then they would represent the first detections of substructure along this axis. The presence of substructure along the minor axis requires that magnetic field variations transverse to the line of sight are not restricted to a narrow angular range (the field wander is more likely to sample all angles, but with a preference for angles that are aligned with the minor axis of the scattering). Minor axis substructure would eliminate, for example, the ``boxcar'' model for refractive fluctuations in \citet{Psaltis_2018}, which describes magnetic field wander as a uniform distribution over a limited range of angles. 

% From these observations, we then draw two primary conclusions related to the scattering. First, the substructure we detect is most likely to be intrinsic, having a stronger signature along the minor axis of the scattering. And, second, models that predict significant substructure along the major axis are ruled out. For instance, the inner 95\% of refractive noise samples fall in the range $[0.16, 1.9] \times \hat{\sigma}$. Thus, the GS06 model significantly over-predicts the refractive noise on all baselines longer than ${\sim}1.8 G\lambda$, even assuming that all the signal seen is from scattering substructure. Thus, we can confidently reject this model for the scattering of \sgra. The J18 model is acceptable, but only if the ALMA-VLBA detections predominantly represent intrinsic structure. This could be tested by examining whether the structure is stable because excess flux density from scattering would vary stochastically among observations separated by more than a few days. 

\begin{figure*}[h!t]
\begin{minipage}[t]{\linewidth}
\includegraphics[width=0.462\textwidth]{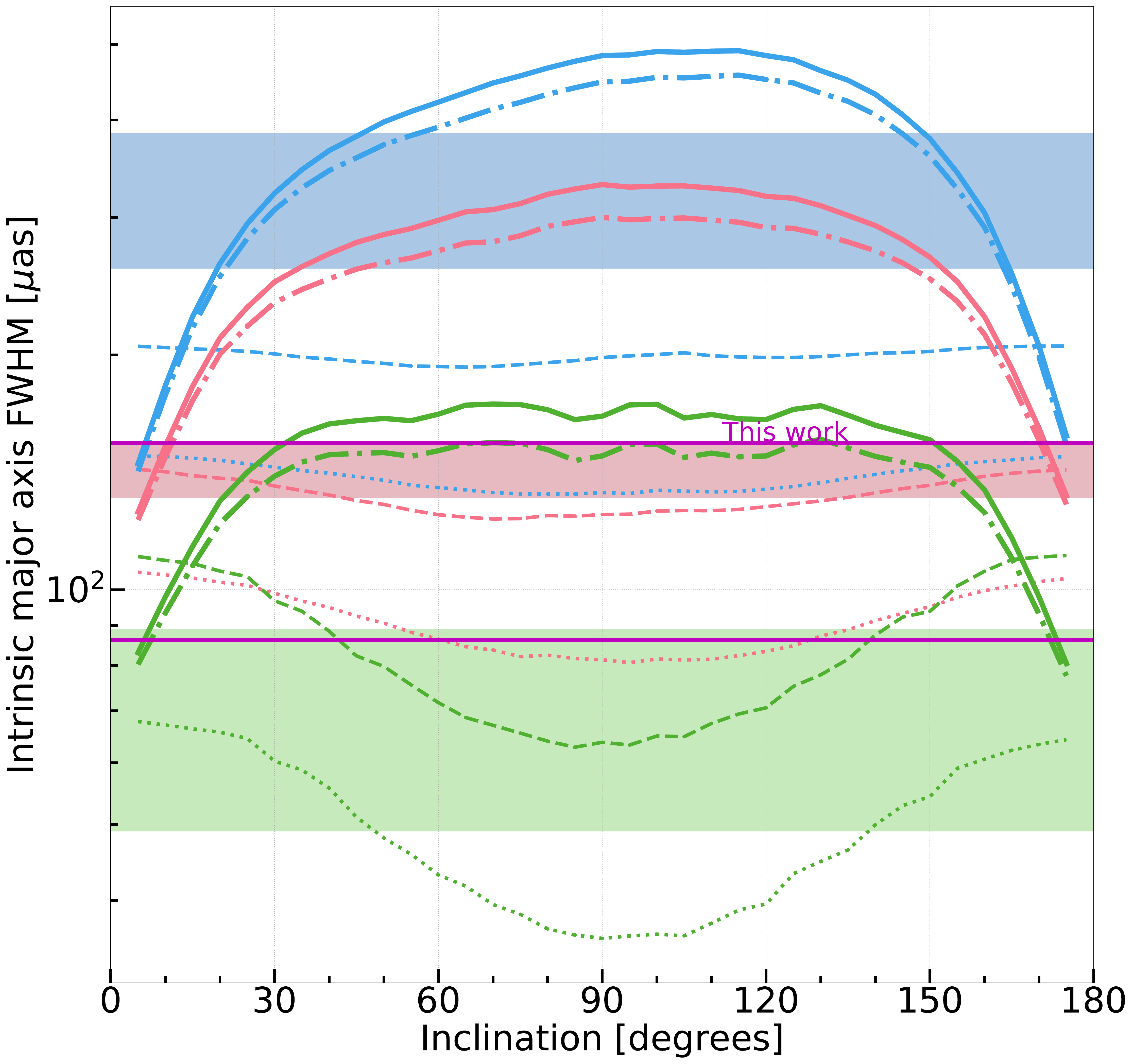}
\includegraphics[width=0.462\textwidth]{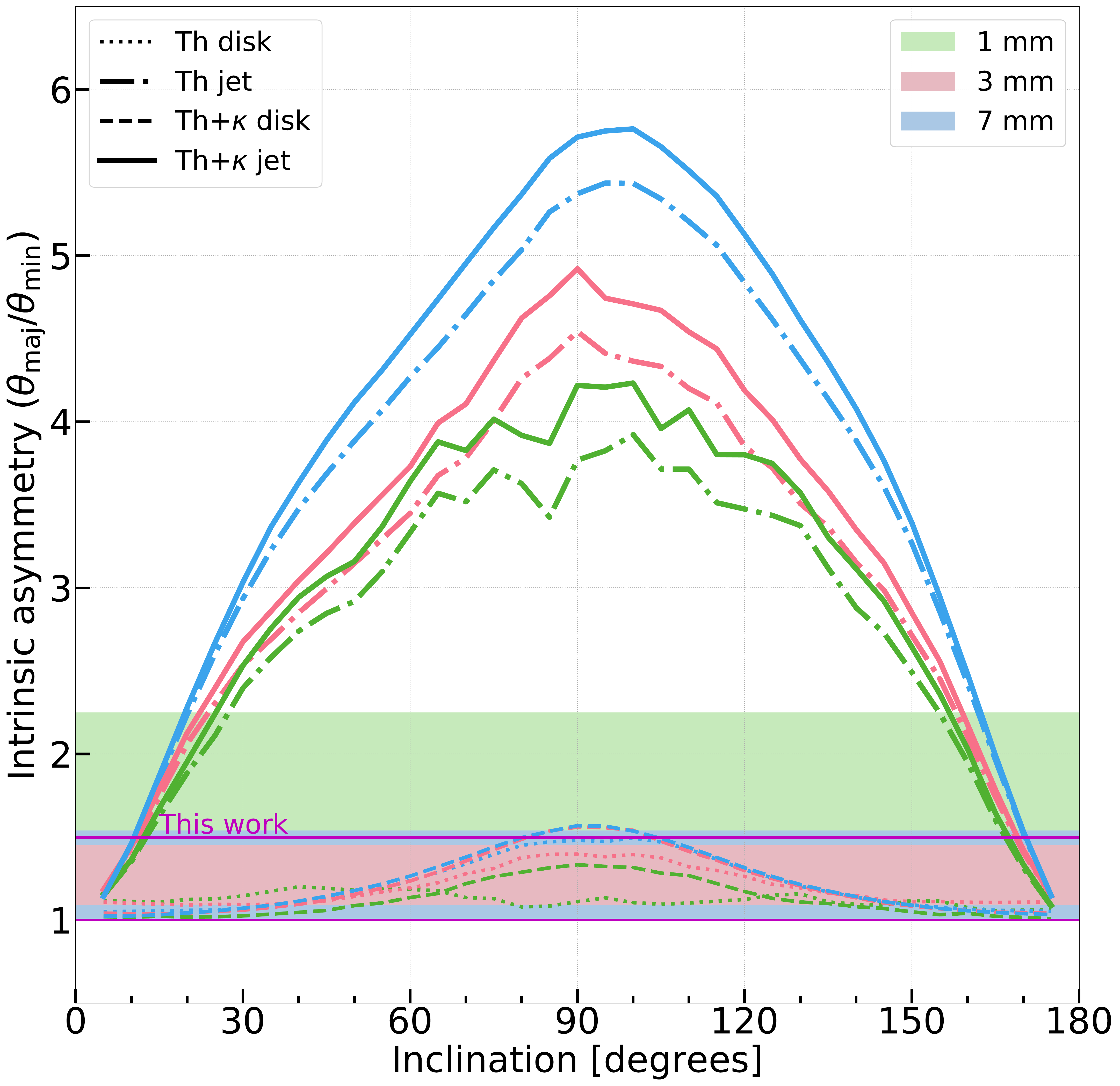}
\caption{Intrinsic size and asymmetry (axial ratio) estimates from observations of \sgra\ at 1, 3 and 7\,mm vs.\ theoretical predictions based on 3D GRMHD simulations of black hole accretion flows. Line color encodes the wavelength of observation and the bands are size and asymmetry bounds from model-fitting \citepalias{Johnson_2018}. The upper and lower size and asymmetry image-domain bounds from this work are shown as solid magenta lines. Data constraints at 1\,mm extend to a lower asymmetry bound of $1.0$. Various line types correspond to models with varying prescriptions for electron acceleration and disk/jet dominated flows generated at each wavelength: ``th'' for a purely thermal disk or jet dominated emission model, ``th+$\kappa$'' for a thermal model with accelerated particles (1\% for disk and 10\% for jet) in a power-law ($\kappa$) distribution \citep{Moscibrodzka_2009,Moscibrodzka_2014,Moscibrodzka_2016,Davelaar_2018}. {\it Left}: Intrinsic source sizes as a function of the viewing angle. {\it Right}: Intrinsic asymmetry (axial ratio) as a function of the viewing angle.}
\label{fig:model_comp}
\end{minipage}
\begin{minipage}[t]{\linewidth}
\includegraphics[width=0.472\textwidth]{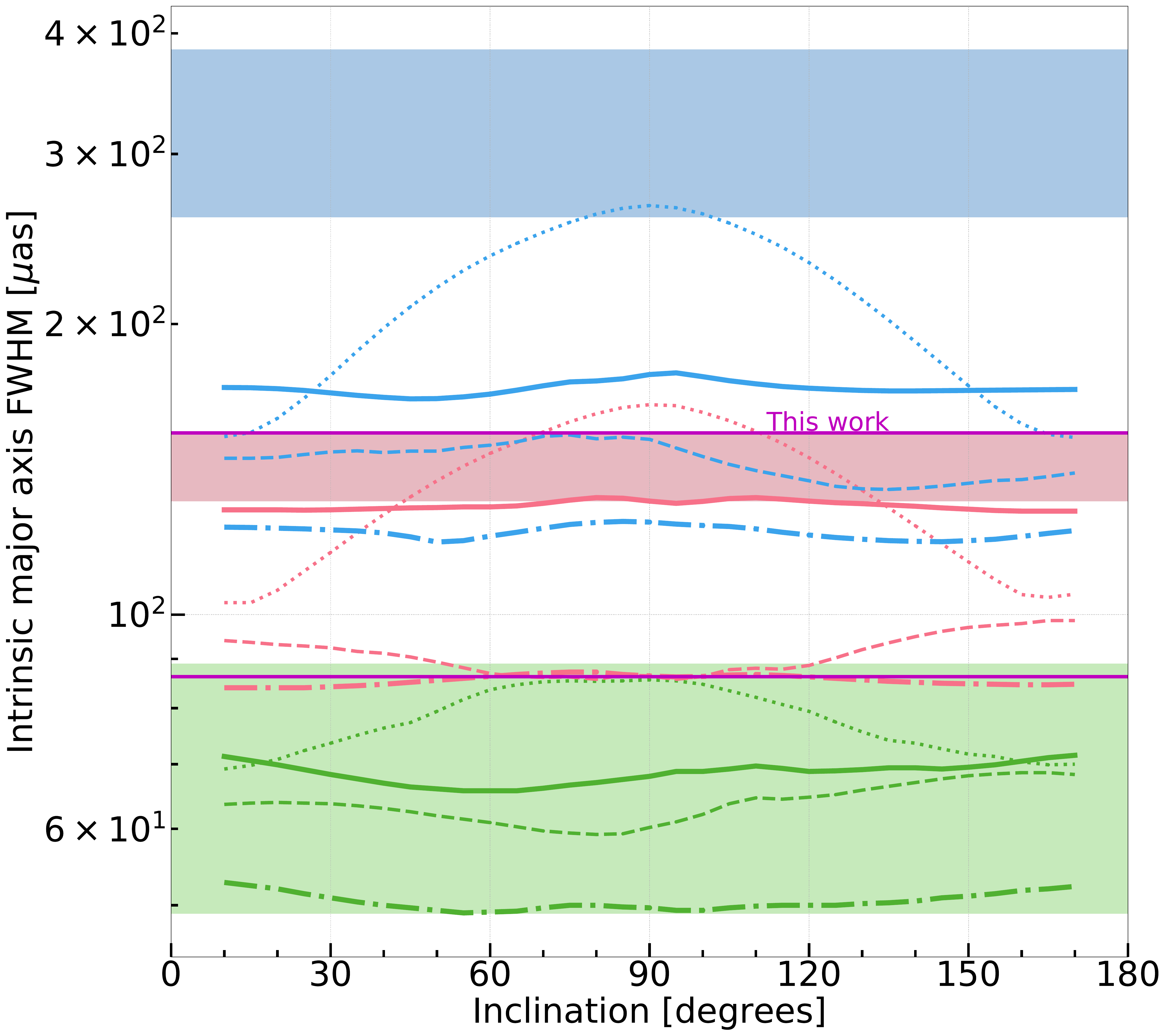}
\includegraphics[width=0.462\textwidth]{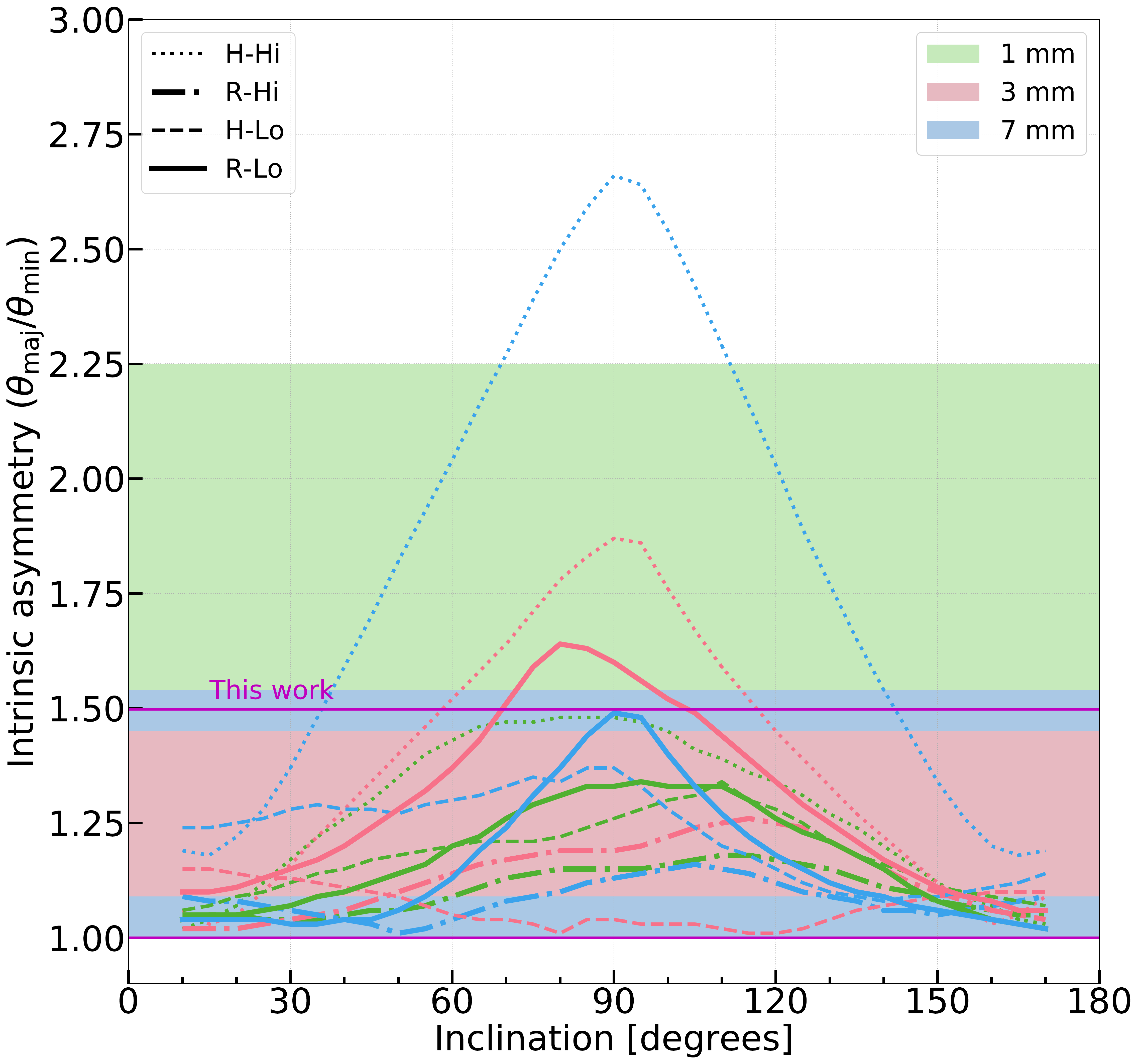}
\caption{Intrinsic size and asymmetry (axial ratio) estimates from observations of \sgra\ at 1, 3 and 7\,mm vs.\ theoretical predictions based on 3D GRMHD simulations of black hole accretion flows. Line color encodes the wavelength of observation and the bands are size and asymmetry bounds from model-fitting \citepalias{Johnson_2018}. The upper and lower size and asymmetry image-domain bounds from this work are shown as solid magenta lines. Data constraints at 1\,mm extend to a lower asymmetry bound of $1.0$. Various line types correspond to models with varying prescription for electron heating and black hole spin generated at each wavelength: ``H'' for the Howes turbulent cascade prescription, ``R'' for the Rowan magnetic reconnection prescription, ``Lo'' for a non-spinning black hole, and ``Hi'' for a black hole with a dimensionless spin of 0.9375 \citep{Howes_2010,Rowan_2017,Chael_2018b}. {\it Left}: Intrinsic source sizes as a function of the viewing angle. {\it Right}: Intrinsic asymmetry (axial ratio) as a function of viewing angle.}
\label{fig:model_comp2}
\end{minipage}
\end{figure*}

\subsection{Constraints on accretion flow and jet models}\label{sec:theory}

The intrinsic image of \sgra\ at 3\,mm shown in Figure~\ref{fig:sgra_so} allows us to discriminate between the two main classes of models that now must fit the tight source size and morphology constraints derived from both model-fitting (from previous experiments) and our image-domain measurements. We can explore a small subset of GRMHD simulations to assess possible constraints from our observables. Due to our unconstrained estimate of the PA, we opted to compare the major axis size and the asymmetry (axial ratio), which are independent of the PA of the source on the sky.

Figure~\ref{fig:model_comp} compares the sizes and morphology of 7/3/1.3\,mm images from a sample of 3D-GRMHD simulations of either disk or jet dominated emission, at varying viewing angle with respect to the black hole spin axis, with observational constraints from current (Table \ref{tab:sgra_size}) and previous observations of \sgra\ \citepalias[see Table 4 in][]{Johnson_2018}. Model images are generated by combining the dynamical model with ray-tracing and radiative transfer using only synchrotron opacities. 
To estimate the size of the radiating region in model images we calculate the eigenvalues of the matrix formed by taking the second central moment of the image on the sky \citep[i.e., the length of the ``principal axes'',][]{Hu_1962}.

 Producing a ray-traced image from single-fluid GRMHD simulations requires providing the electron distribution function (hereafter eDF), which is unconstrained in traditional single-fluid GRMHD simulations. Thermal disk models (``Th disk'' in Figure~\ref{fig:model_comp}) assume a thermal, Maxwell-J{\"u}ttner eDF and a proton-to-electron temperature ratio\footnote{Standard GRMHD simulations provide only the fluid pressure, which is dominated by the protons. In a perfect fluid, the pressure in a grid zone gives a proton temperature. For strongly sub-Eddington accretion flows with $L_{\rm Bol}/L_{\rm Edd} \approx 10^{-8}$, protons and electrons are not necessarily well coupled by Coulomb collisions. In these GRMHD simulations the electron temperatures are not self-consistently computed but they are essential in calculating synchrotron emission. The electron temperature is parameterized by a coupling ratio, $T_{\rm p}/T_{\rm e}$, between the proton and electron temperature.} $T_{\rm p}/T_{\rm e}=3$ everywhere \citep[motivated by results of ][]{Moscibrodzka_2009}. Models denoted as ``Th jet'' have $T_{\rm p}/T_{\rm e}=20$ in the accretion disk and $T_{\rm p}/T_{\rm e}=1$ along the magnetized jet, which allows the jet to outshine the disk at mm-wavelengths (this jet model has been introduced by \cite{Moscibrodzka_2014,Moscibrodzka_2016}). There is a family of models in-between these two extreme cases. 
In the models denoted as ``Th+$\kappa$  disk'' the eDF is hybrid; 1 percent of all electrons are non-thermal, described by a $\kappa$ eDF. Adding non-thermal electrons to the emission model results in more extended disk images as the non-thermal electrons produce a diffuse ``halo'' around the synchrotron photosphere. The ``halo'' contributes to the disk size estimates \citep{Mao_2017}. Finally, the ``Th+$\kappa$ jet'' model is a 3D version of the $\kappa-$jet model introduced by \cite{Davelaar_2018} with 10 percent of jet electrons in a $\kappa$ eDF. In both hybrid models the $\kappa$ parameter is set to 4 \citep[see][for details]{Davelaar_2018}.

We find that only disks with a hybrid eDF at moderate viewing angles and both jet-models with viewing angles $\lsim{20^\circ}$ are consistent with 1.3 and 3\,mm sizes and asymmetry constraints. This limit is consistent with the recent low-inclination constraints derived from orbital motions in near-infrared \sgra\, flares by \citet{Gravity_2018b} observed with the GRAVITY instrument. In the tested models, the dependency of the source sizes as a function of observing wavelength is shallower than the $\theta \sim \lambda$ dependency estimated from multi-wavelength observations of \sgra\ \citepalias[Figure 13 in][]{Johnson_2018}. Hence none of the models that satisfy 1.3/3\,mm source sizes can account for the 7\,mm source size.

Although GRMHD simulations of black hole accretion are inherently time-variable, causing the size and asymmetry to fluctuate in time, these changes are smaller than 10 percent. We conclude that current models under-predict the observed 7\,mm emission size, even when accounting for size and asymmetry fluctuations in time. In simulations, the 7\,mm photons are emitted from larger radii where the accretion flow structure is less certain due to lower grid resolution, the initial conditions (finite size torus with pressure maximum at $r=24~GM/c^2$) and boundary conditions of the simulation that only allow for plasma outflows. These issues as well as the electron acceleration should be addressed by future radiative GRMHD simulations of \sgra.

We also explored another set of 3D simulations from \citet{Chael_2018b}, performed with the two-temperature, radiative GRMHD code \texttt{KORAL}
\citep[][see Figure~\ref{fig:model_comp2}]{Sadowski_2013,Sadowski_2014,Sadowski_2017}. Unlike the simulations presented in Figure~\ref{fig:model_comp}, where the electron temperature (and potential non-thermal component) is assigned to the simulation in post-processing, \texttt{KORAL} evolves the electron temperature throughout the simulation self-consistently with contributions from radiative cooling, Coulomb coupling, and dissipative heating. While the physics of radiation and Coulomb coupling is well understood, the dissipative heating of electrons and ions is governed by unconstrained plasma microphysics that occurs at scales far smaller than the grid scale of the simulation. 

\citet{Chael_2018b} investigated two different physical prescriptions for the electron dissipative heating. The first prescription is the Landau-damped turbulent cascade model of \citet{Howes_2010}. Since this prescription primarily heats electrons in regions where the plasma is highly magnetized, it produces prominent emission from the jet and outflow of the GRMHD simulations at 3.5\,mm\citep[see also][]{Ressler_2017}. The other prescription for electron heating investigated in \citet{Chael_2018b} is based on particle-in-cell simulations of particle heating from magnetic reconnection presented in \citet{Rowan_2017}. This prescription heats electrons and ions equally and only in highly magnetized regions, resulting in cooler jet regions with less emission than the disk. In total, \citet{Chael_2018b} presented four simulations spanning the two heating prescriptions considered (``Howes'' or ``H'' for the turbulent cascade prescription of \citealt{Howes_2010} and ``Rowan'' or ``R'' for the reconnection prescription of \citealt{Rowan_2017}) and two values of the dimensionless black hole spin ($a=0$ for ``Lo'', and $a=0.9375$ for ``Hi'').

Figure~\ref{fig:model_comp2} shows that all four models presented in \citet{Chael_2018b} fit the 1.3\,mm constraints and mostly fit the 3\,mm image-domain constraints. However, only the H-Hi and R-Lo models fit the model-fitting 3\,mm range at moderate viewing angles, and all models fail to match 7\,mm constraints. However, these simulations were only run over a relatively short time, and inflow equilibrium in the disk was only established up to $\sim$20 $R_{\rm Sch}$, while the 7\,mm emission extends to $\sim$35 $R_{\rm Sch}$. To accurately compare the predictions from these two heating models with predictions at 7\,mm and longer wavelengths, the simulations will have to be run longer using initial conditions adapted to producing an accretion disk in equilibrium past 20 $R_{\rm Sch}$.

\begin{figure}[t]
\includegraphics[width=\linewidth]{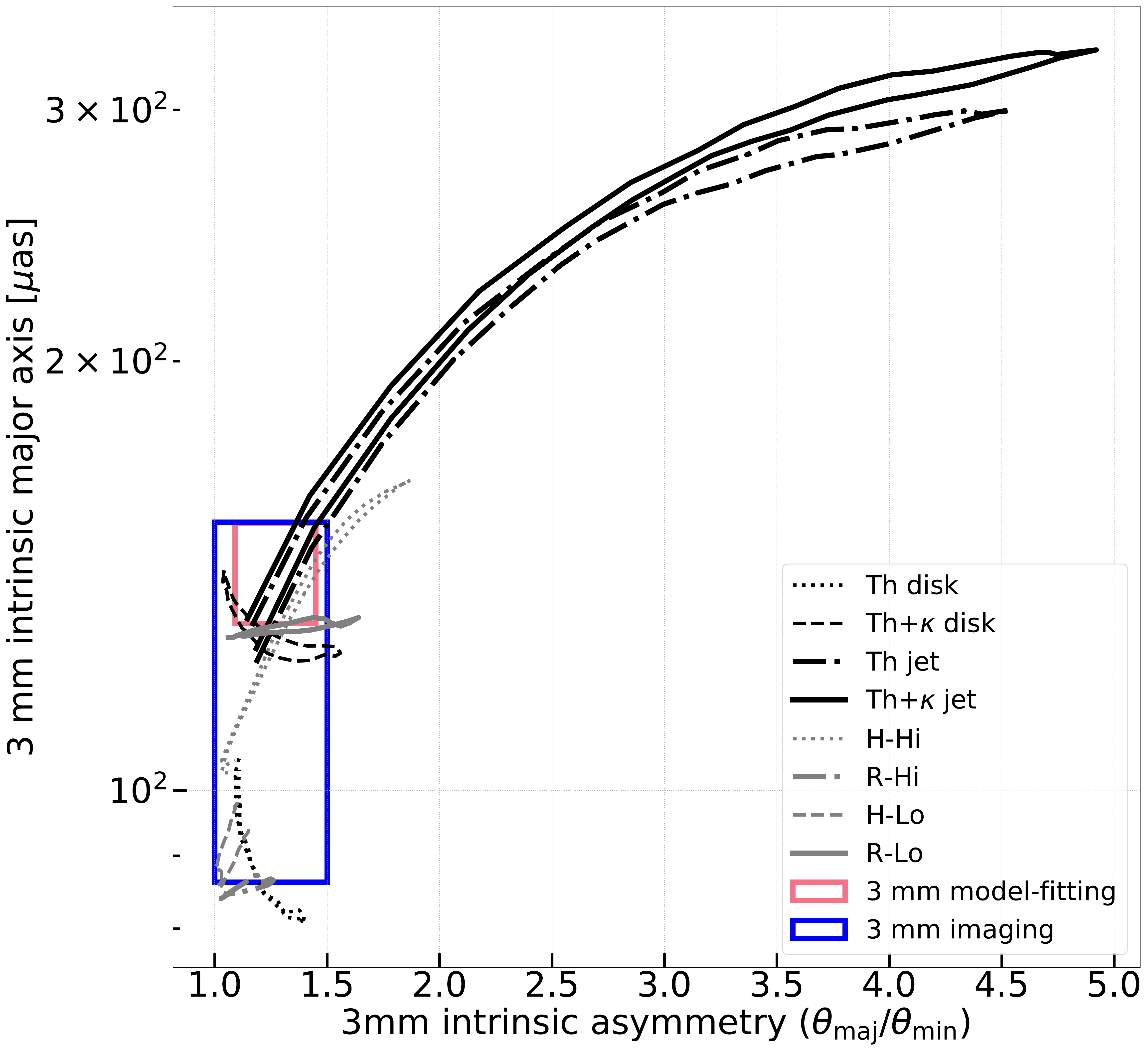}
\caption{3\,mm models compared to the plausible range from 3\,mm data via model-fitting \citepalias{Johnson_2018} and image-domain constraints (this work). Various line types correspond to models with varying prescriptions for electron acceleration/heating \citep{Moscibrodzka_2009,Moscibrodzka_2014,Moscibrodzka_2016,Davelaar_2018,Chael_2018b}.}
\label{fig:model_3mm}
\end{figure}

\begin{figure}[h!t]
\centering 
\includegraphics[width=0.95\linewidth]{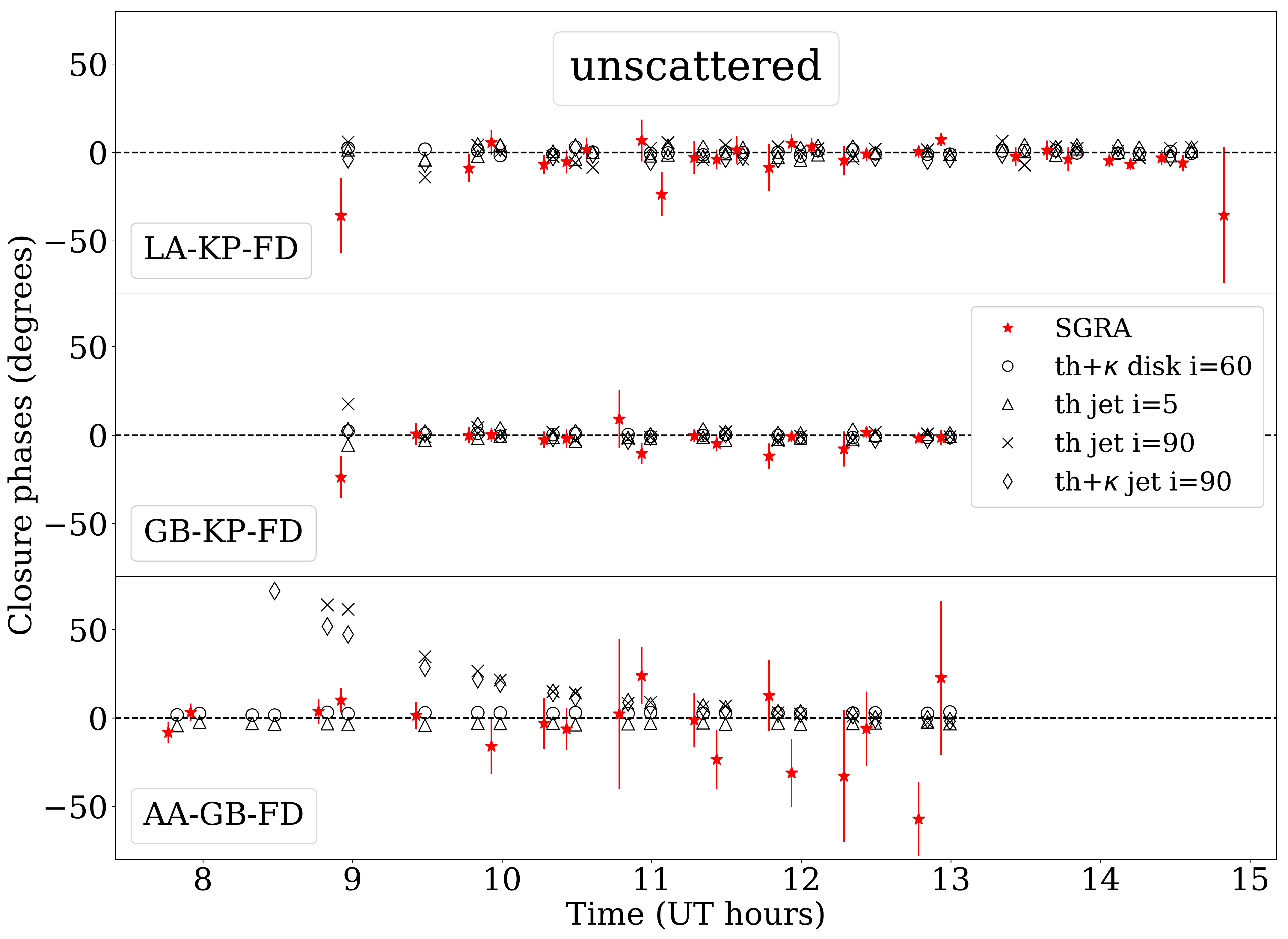}
\includegraphics[width=0.95\linewidth]{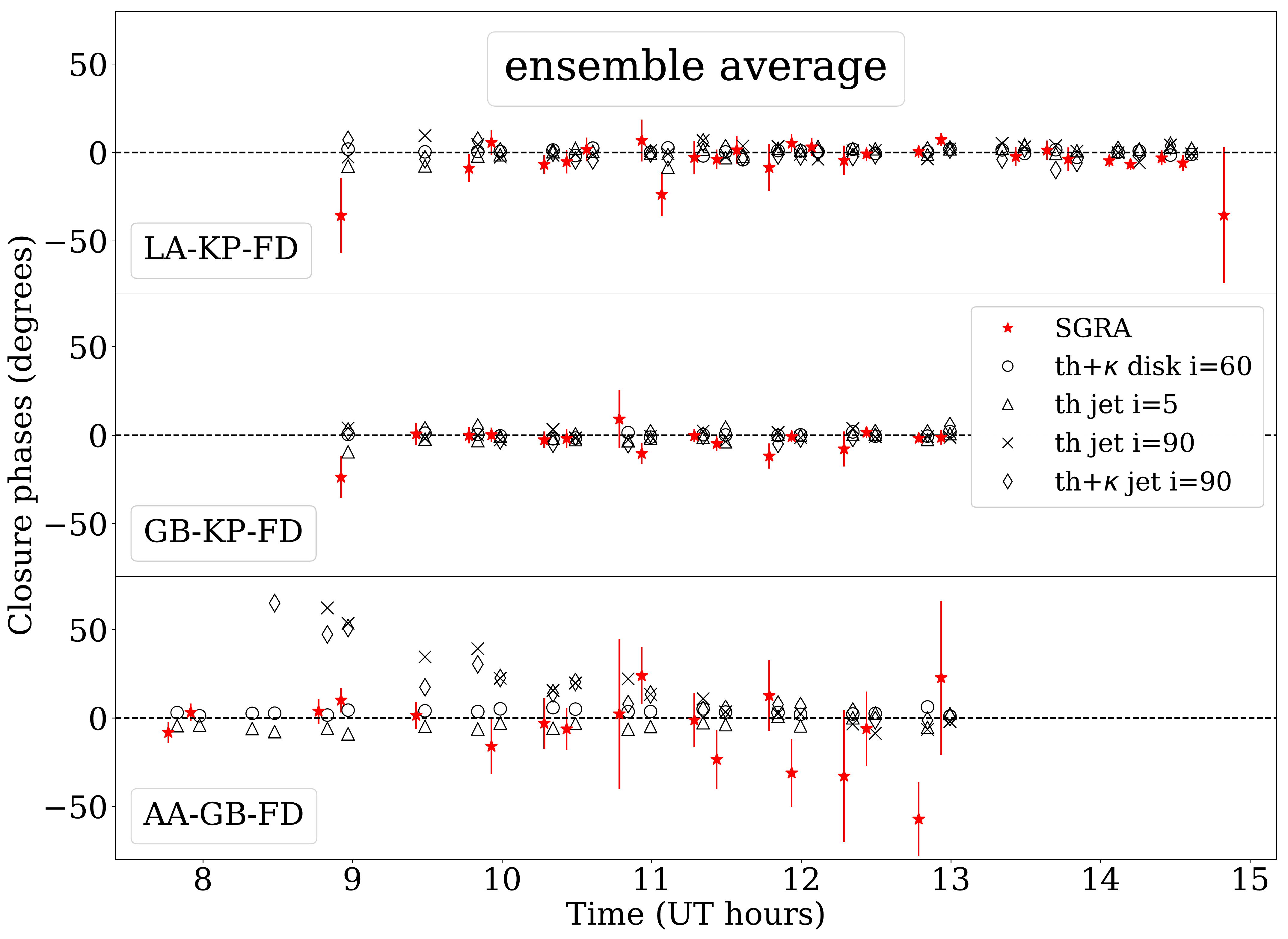}
\includegraphics[width=0.95\linewidth]{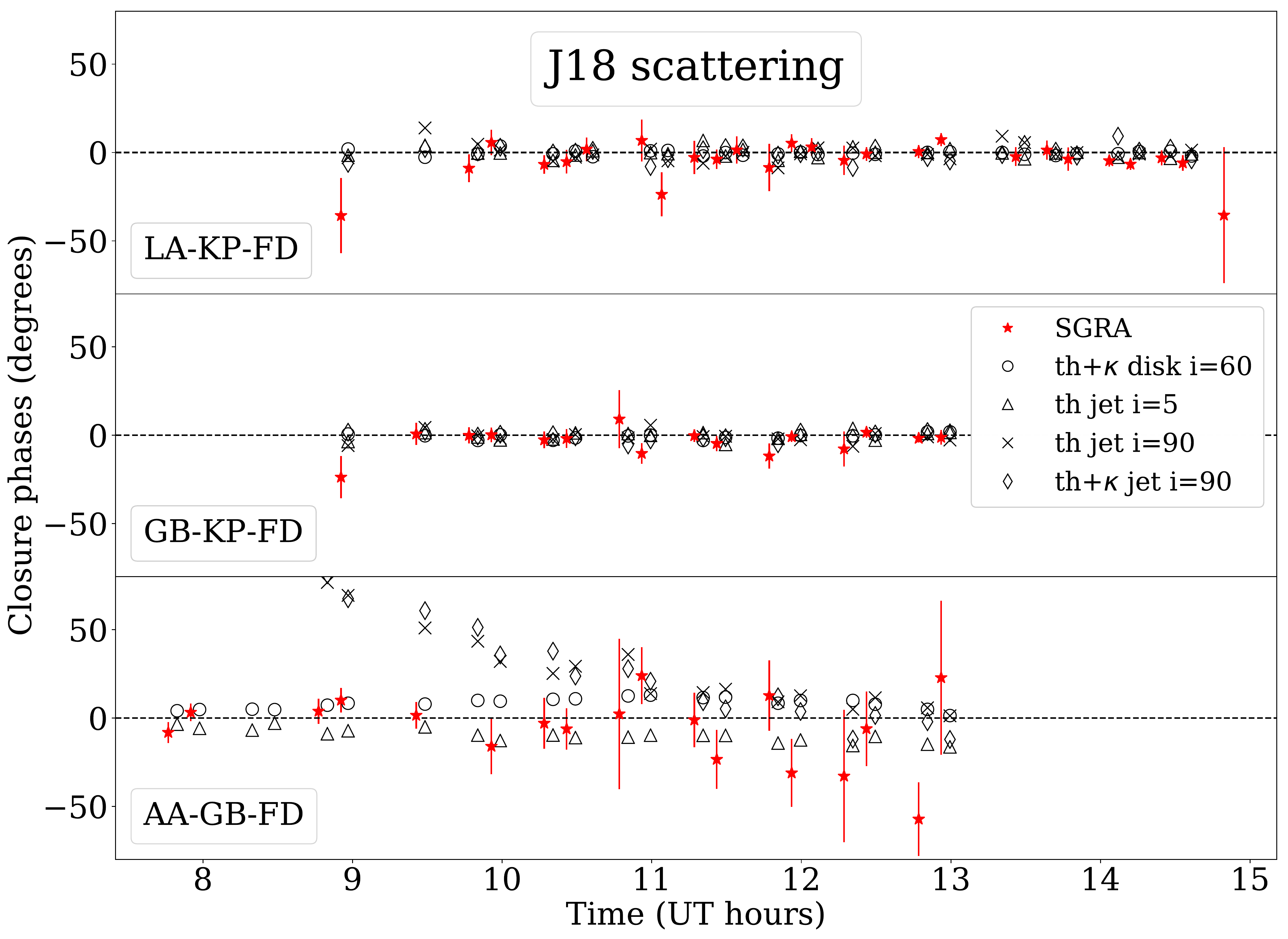}
\caption{Scan-averaged closure phases for \sgra\ on three triangles (LA-KP-FD, GBT-KP-FD, ALMA-GBT-FD) with predictions for a thermal$+\kappa$-distribution disk model (th$+\kappa$ disk, $i=60^\circ$), thermal$+\kappa$-distribution jet model (th$+\kappa$ jet, $i=90^\circ$) and thermal jet model (th jet, $i=5^\circ$ and $i=90^\circ$), where $i$ is the inclination. Each model is shown without scattering (top), with ensemble-average scattering (center), and with a single realization of scattering (bottom). Note that ensemble-average scattering does not affect closure phase, and even a single realization of the scattering has little effect on the closure phases for these triangles.
}
\label{fig:model_cphase}
\end{figure}

Figure~\ref{fig:model_3mm} demonstrates the plausible range of intrinsic 
source sizes vs.\ asymmetries at 3\,mm for all of the models we have explored. Here it is evident which models fall into the permitted region. 
Given that our modeling does not involve any detailed parameter fitting, the agreement between models and observables is encouraging. Disk and jet models with different heating prescriptions are also likely to have distinct polarimetric characteristics that can be compared to observables \citep[e.g.,][]{Gold_2017,Moscibrodzka_2017}. 

Furthermore we can directly compare closure phases from the different models with those presented in Section~\ref{sec:results}. Closure phases observed are an additional robust criterion to discriminate between models: they are independent of imaging assumptions, the beam of the observations, and scattering effects. In Figure~\ref{fig:model_cphase} we compare the scan-averaged closure phases from \sgra\ for the three representative triangles to four example models: the thermal+$\kappa$ disk model at an inclination of 60$^\circ$ and the thermal jet model at an inclination of 5$^\circ$, which fit the major axis and asymmetry bounds given by the 3.5\,mm observations; and the thermal+$\kappa$ jet model and thermal-only jet model, both at an inclination of 90$^\circ$, which do not fit the 3.5\,mm bounds. We simulated observations of the four different models with the same stations and coverage as our GMVA+ALMA dataset and compare the closure phases for the original model images (Figure~\ref{fig:model_cphase}; top panel), for the model scattered with the J18 refractive scattering (center panel), and for the ``ensemble average'' models scattered only with the scattering kernel (no refractive noise; bottom panel).
We find that for the small and medium triangles it is very difficult to distinguish between models as they all have closure phases near zero, similar to our measurements \citep{Fraga_2016}. However, for the large triangle (ALMA-GBT-FD), two models show strong non-zero closure phases: the thermal+$\kappa$ and the thermal-only jet models at 90$^\circ$ inclination. Interestingly, these are also the example models that do not fit the intrinsic asymmetry and size bounds from 3.5\,mm. 
We also find that interstellar scattering as modeled by \citetalias{Johnson_2018} does not strongly affect intrinsic closure phase: for both the ensemble average and fully scattered cases, the two jet models at 90$^\circ$ inclination clearly deviate from what is measured on the largest triangle. The Howes and Rowan models are omitted from the comparisons in Figure~\ref{fig:model_cphase} for clarity, as they are all very symmetrical and compact: their closure phase behavior is similar to the thermal+$\kappa$ disk and the thermal jet models nearly or fully pointed along the line of sight. 

While our comparisons to simulations are limited to a handful of GRMHD models, they demonstrate the strong constraints provided by multi-wavelength measurements of size, shape and point-symmetry of \sgra.

\section{Summary}\label{sec:summary}
%{\color{red} Lynn: Parts of the last paragraph of Section 2 are very important for explaining why 3mm observations are uniquely valuable. Perhaps move some of this to the introduction and/or consider re-iterating in the summary? In other words, be sure that even a casual reader understands that this study will not superseded by work at 1mm.} {\color{blue} I think Lynn means here to reiterate why emission at 3mm is more valuable to study accretion physics, versus 1mm being dominated by relativistic effects?} 

We have presented observations of \sgra\ using ALMA in concert with the GMVA at 86\,GHz. These are the first observations to use ALMA as part of a VLBI array, improving the angular resolution for observations of \sgra\ at this frequency by more than a factor of two. The improved resolution and sensitivity have allowed us to reconstruct an intrinsic image of \sgra\ for the first time at this frequency, which is also the first image of \sgra\ for which the scattering is subdominant to intrinsic structure. We find that the intrinsic image of \sgra\ has an asymmetry (axial ratio) of $1.2^{+0.3}_{-0.2}$ and a major axis of $120\pm34~\mu$as, although we cannot constrain the position angle because of the highly symmetric intrinsic source.

We have demonstrated that the geometrical properties of the intrinsic image and observed closure phases tightly constrain accretion flow models onto \sgra. Our measurements require models to have symmetrical morphology, 86\,GHz radio emission spanning $12 \pm 3.4$ Schwarzschild radii, and closure phases close to zero on the triangles sampled in our observation. For the eight theoretical simulations we have considered at 3.5\,mm, our data are compatible with disk models at all inclinations and jet models fully or nearly pointed along the line of sight. None of the simulations we consider is able to simultaneously match size and asymmetry limits from 1, 3, and 7\,mm observations due to the relatively small domain simulated by state of the art 3D GRMHD models.

While GRMHD models are promising to describe emission near the horizon, semi-analytical models for the accretion flow and jet can be more readily extended to larger domains \citep[e.g.,][]{Broderick_2016,Gold_2017,Pu_2018}. In addition, it is more straightforward to explore parameter dependencies for semi-analytic models. The model of \citet{Broderick_2016} is compatible with our 3.5\,mm size and asymmetry estimates. Exploring whether these models can be compatible with the full set of multi-wavelength size and asymmetry constraints for \sgra\ is a promising avenue for continued study.

In addition to the overall image morphology, we have discovered non-Gaussian structure along the minor axis of \sgra, hinting at either a non-Gaussian intrinsic source or a non-Gaussian scattering kernel. Comparisons of the observed visibility amplitudes against two scattering models showed that the scattering model presented by \cite{Goldreich_2006} over-predicts the correlated flux density on long baselines to ALMA (1.8-2.4~G$\lambda$). This model also overpredicts the flux density on east-west baselines longer than 1~G$\lambda$. Thus, this model for the scattering of \sgra\ is conclusively ruled out by our observations. The exclusion of the GS06 model shows that refractive scattering is likely to weakly affect 1.3\,mm images with the EHT. 

The scattering model presented by \citet{Johnson_2018}, on the other hand, predicts comparable levels of refractive noise to the excess flux density we have observed on baselines above 1.8~G$\lambda$. However, using our single observation with ALMA, we cannot conclusively determine whether those detections are entirely due to refractive noise or if they are a combination of intrinsic source structure and scattering substructure. Continued observations of \sgra\ will elucidate these questions, including deeper VLBI observations at 22 and 43\,GHz to better estimate the inner scale from the shape of the scatter-broadening kernel \citep[e.g.,][]{Zhao_2018}, and additional GMVA+ALMA observations that will sample different realizations of the scattering screen. 

\acknowledgements{This work is supported by the ERC Synergy Grant ``BlackHoleCam: Imaging the Event Horizon of Black Holes'', Grant 610058. We thank the National Science Foundation (AST-1126433, AST-1614868, AST-1716536) and the Gordon and Betty Moore Foundation (GBMF-5278) for financial support of this work. This work was supported in part by the Black Hole Initiative at Harvard University, which is supported by a grant from the John Templeton Foundation. K.~A. is a Jansky Fellow of the National Radio Astronomy Observatory. M.~K. acknowledges the financial support of JSPS KAKENHI grants No. JP18K03656 and JP18H03721. R.-S. L. is supported by the National Youth Thousand Talents Program of China and by the Max-Planck Partner Group. L.~L. acknowledges the financial support of DGAPA, UNAM (project IN112417), and CONACyT, M{\'e}xico. I.~C. acknowledges the financial support of the National Research Foundation of Korea (NRF) via a Global PhD Fellowship Grant (NRF-2015H1A2A1033752).  This paper makes use of the following ALMA data: ADS/JAO.ALMA2016.1.00413.V. ALMA is a partnership of ESO (representing its member states), NSF (USA) and NINS (Japan), together with NRC (Canada), MOST and ASIAA (Taiwan), and KASI (Republic of Korea), in cooperation with the Republic of Chile. The Joint ALMA Observatory is operated by ESO, AUI/NRAO and NAOJ. This research has made use of data obtained with the Global Millimeter VLBI Array (GMVA), which consists of telescopes operated by the (Max-Planck-Institut f{\"u}r Radioastronomie) (MPIfR), IRAM, Onsala, Metsahovi, Yebes, the Korean VLBI Network, the Green Bank Observatory and the Long Baseline Observatory (LBO). The VLBA is an instrument of the LBO, which is a facility of the National Science Foundation operated by Associated Universities, Inc. The data were correlated at the DiFX correlator of the MPIfR in Bonn, Germany. This work is partly based on observations with the 100-m telescope of the MPIfR at Effelsberg. This work made use of the Swinburne University of Technology software correlator \citep{Deller_2011}, developed as part of the Australian Major National Research Facilities Programme and operated under licence.}

\software{
AIPS \citep{Greisen_2003},
DiFX \citep{Deller_2011},
HOPS \citep{Whitney_2004,lindyhops},
KORAL \citep{Sadowski_2013,Sadowski_2014,Sadowski_2017},
PolConvert \citep{Marti_2016},
eht-imaging library \citep{Chael_2016},
Stochastic Optics \citep{Johnson_2016}
}

\bibliography{3mm_bib.bib}

\end{document}